\documentclass[structabstract]{aa}
\usepackage{graphicx}
\usepackage{lscape}
\def\approxgt{\mathrel{\hbox{\rlap{\lower.55ex \hbox {$\sim$}}
        \kern-.3em \raise.4ex \hbox{$>$}}}}
\def\approxlt{\mathrel{\hbox{\rlap{\lower.55ex \hbox {$\sim$}}
        \kern-.3em \raise.4ex \hbox{$<$}}}}
\begin{document}
   \title{On the driver of relativistic effects strength in Seyfert galaxies}

   \author{
          M.~Guainazzi
          \inst{1},
	  S.~Bianchi
          \inst{2,3},
	  I.~de la Calle P\'erez
          \inst{1},
	  M.~Dov\v ciak
	  \inst{4},
	  A.L.Longinotti
          \inst{5}
          }

   \offprints{M.Guainazzi}

   \institute{
              $^1$European Space Astronomy Centre of ESA P.O.Box 78,
              Villanueva de la Ca\~nada, E-28691 Madrid, Spain \\
	      \email{Matteo.Guainazzi@sciops.esa.int} \\
	      $^2$Dipartimento di Fisica, Universit\`a degli Studi Roma Tre,
	      via della Vasca Navale 84, I-00146 Roma, Italy \\
	      $^3$INAF-Osservatorio Astronomico di Brera, Via E.Bianchi 46,
	      I-23807, Merate, Italy \\
	      $^4$Astronomical Institute, Academy of Sciences, Bo\v cn\'i II,
	      14131 Prague, Czech Republic \\
	      $^5$MIT Kavli Institute for Astrophysics and Space Research,
	      77 Massachusetts Avenue, NE80-6011, Cambridge, MA, 02139
              }

   \date{Received ; accepted }

   \abstract{Spectroscopy of X-ray emission lines emitted in accretion discs around supermassive black holes is
             one of the most powerful probes of the accretion flow physics and geometry, while also providing
             in principle observational constraints on the black hole spin. Previous studies have suggested that
             relativistically broadened line profiles are fairly common in nearby unobscured
             Seyfert galaxies. Their strength, as parametrised by the Equivalent Width (EW) against the total
             underlying continuum, spans a range of almost two orders of magnitude.}
	    {We aim at determining the ultimate physical driver of the strength of this relativistic reprocessing
	     feature.}
	    {We first extend the hard X-ray flux-limited
	     sample of Seyfert galaxies studied so far (FERO,
	     de la Calle P\'erez et al. 2010) to obscured objects up to a column density
	     $N_H$=$6 \times 10^{23}$~cm$^{-2}$.
	     We verify that none of the line
	     properties depends on the AGN optical classification,
	     as expected from the Seyfert unification
	     scenarios. There is also no correlation between the accretion disc inclination, as derived from
	     formal fits of the line profiles, and the optical type or host galaxy aspect angle,
	     suggesting that
	     the innermost regions of the accretion disc and the host galaxy plane are not aligned.
	     We use this extended sample to study the EW dependency on various observables,
	     and compare it with the predictions of Monte-Carlo accretion disc reprocessing simulations
	     (George \& Fabian 1991).}
	    {The behaviour of the EW as a function
	     of disc inclination, shape of the intrinsic power-law nuclear continuum, or
	     iron abundance does not agree with the simulation predictions. Data are not sensitive enough to
	     the detailed ionisation state of the line-emitting disc. However, the lack of dependency of
	     the line EW on either the luminosity or the rest-frame centroid energy
	     rules out that disc ionisation plays an important role on the EW dynamical range in
	     Seyferts.}
	    {The dynamical range of the relativistically broadened K$_{\alpha}$ iron
	     line EW in nearby Seyferts appears to be mainly determined by the properties of the innermost accretion flow.
	     We discuss several mechanisms (disc ionisation, disc truncation, aberration due to a mildly
	     relativistic outflowing corona) which can explain this. We stress that the above results as such do
	     not represent either a falsification or a proof the relativistically blurring scenario. Observational data
	     are still not in contradiction with scenarios invoking different mechanisms for the spectral
	     complexity around the iron line, most notably the ``partial covering'' absorption scenario.}


\authorrunning{Guainazzi et al.}

\titlerunning{The ultimate driver of relativistic effects in AGN}

\maketitle
%

\section{X-ray spectroscopic evidence for relativistic accretion flow in AGN}

Accretion discs feeding black holes are predicted to dissipate most of their energy in the innermost few Schwarzschild
radii (\cite{agol00,krolik02}). In these regions, the space-time distortion affects significantly the photon path
through general and special relativity effects (\cite{fabian00,reynolds03}). Monochromatic lines emitted in relativistic
discs are expected to suffer significant profile distortions (\cite{fabian89,laor91}). The properties of the profiles as
seen at infinity depend on a number of parameters related to the accretion flow (size and ionisation state of the
disc photosphere), its orientation with respect to the line of sight, the shape of the illuminating ionising
continuum (\cite{matt92,zycki94,nayakshin00,ross05}) and the metal abundances. The profile in principle
probes the physical and geometrical properties of the accretion flow on spatial scales inaccessible by any other
conceivable experiments in the electromagnetic domain.

That's the reason why great excitement followed the first detection of a broadened and skewed profile 
of the iron K$_{\alpha}$ fluorescent iron line by ASCA in MCG-6-30-15, a nearby X-ray bright Seyfert~1 galaxy
(\cite{tanaka95}). Broad iron lines were common in ASCA spectra of AGN (\cite{nandra97}). With the launch of major X-ray
observatories at the beginning of the century - {\it Chandra} and XMM-Newton - the hope arose to be able to
accumulate large samples of broad lines, which may unveil experimental clues on the
way matter accretes onto supermassive black holes, as well as allow us to study its dependence
on the AGN and host galaxy properties. This hope has been only partly fulfilled. Although there are plenty of
measurements of broad iron K$_{\alpha}$ lines (\cite{miller07})
(as well as from other elements or iron transitions: \cite{mason03,fabian09}),
many of them have been challenged as non-unique solutions to the complexity of the AGN spectra
(\cite{reeves04,turner09,miller10a}). The controversy is fierce. This controversy will {\it not} be the subject
of this paper. We will assume in the following that relativistic broadening and skewing
is the correct astrophysical scenario to interpret the observed line profiles. If this assumption turns out
to be wrong, the conclusions of this paper will be, it goes without saying, wrong as well.

Recently studies of large AGN samples focusing on the properties of their relativistic lines have
been published (\cite{nandra07,delacalle10}). 
The main conclusions of these studies can be summarised as follows:

\begin{itemize}

\item relativistically broadened iron K$_{\alpha}$ fluorescent lines are present in 35--45\% of bright
nearby ($z \le 0.01$)
Seyfert galaxies. However, taking properly into account the sensitivity limits of the available 
observations, the presence of a broad line with $EW >$30~eV cannot be excluded in 87\% of the parent population.
At even lower EW values one hits the parameter space where systematic calibration
uncertainties dominate

\item the line shape is consistent with it being emitted in a moderately relativistic accretion disc,
as parametrised by the power-law index of the emissivity radial dependence, $\alpha$ ($\langle \alpha \rangle =
2.4 \pm 0.5$) once emission down to the innermost circular stable orbit is assumed. In a few cases, however,
data require a significantly steeper law (notably MCG-6-30-15, \cite{wilms01,fabian03}; 1H0707-495,
\cite{fabian09}; IRAS13224+3803, \cite{ponti10}).
This has been interpreted as evidence in favour of extreme light bending affecting
the line photon paths (\cite{miniutti04})

\item the accretion disc inclination distribution, $\imath$, is peaked around $\simeq$30$^{\circ}$. However, this could be
primarily due to an observational bias against lines produced in highly inclined discs (\cite{matt92}).
Indeed, the measured disc inclination distribution is still consistent with an intrinsic random distribution
(\cite{nandra97,nandra07,delacalle10}).

\item the line Equivalent Width (EW) spans a range of almost one order of magnitude ($\sim$30--300~eV).
In a few cases, even larger EW have been measured (4U1344-60, \cite{piconcelli06}; AXJ0447-0627, \cite{dellaceca05}:
H1413+117, \cite{chartas07}; IRAS~1334+2438, \cite{longinotti03}:
MCG-02-14-009, \cite{porquet06}; Mkn~335, \cite{longinotti07}; NGC~1365, \cite{risaliti09};
PG1543+489, \cite{vignali08}; RBS1423, \cite{krumpe07}), again possibly indicative of reflection-dominated
states induced by strong light bending (\cite{miniutti04})

\item the line properties and intensities are independent of the black hole mass (as the theory postulates;
\cite{fabian89,matt92}), luminosity or accretion rate

\end{itemize}

A standard plane-parallel X-ray illuminated accretion disc covering a 2$\pi$ solid angle should produce an
EW$\sim$150~eV for a typical AGN X-ray spectral shape and solar abundances (\cite{george91}).
The less-than-100\% detection fraction,
as well as this large variety of EW values need an explanation.

In this paper we aim at investigating the ultimate physical
driver of the observed broad iron line strength. For this purpose we first extend the samples published so
far, which included primarily X-ray unobscured AGN, to the whole Seyfert population up to column densities whose
photoelectric cut-off does not hamper the measurement of the continuum underneath the line (Sect.~2).
This allow us to
investigate the correlation between the detection of a relativistically broadened iron line and AGN
classification and host galaxy properties. Subsequently (Sect.~3) we perform a falsification test of the
currently accepted paradigm that the main driver of the line EW is the
solid angle covered by the disc at the X-ray source,
possibly modified by General Relativity effects for the strongest
lines. We discuss our results in Sect.~4.

\section{The GREDOS sample}

We extract  the GREDOS ({\it General Relativity Effects Detected in Obscured Sources}) sample
by correlating the 54 months 14-195~keV 
{\it Swift} {\it Palermo BAT Catalogue} (PBC; Cusumano et al. 2010) with the XMM-Newton observation
log as of August 2010.
The parent sample includes 754 AGN.
The definition of the GREDOS sample is solely based on X-ray spectral properties.
In order for a source to to belong to GREDOS, the column density measured from a simple power-law continuum fit
of the XMM-Newton EPIC spectra
in an optimal
observation-dependent
energy band has to be comprised between
$5 \times 10^{21}$
and $6 \times 10^{23}$~cm$^{-2}$. 
Actually, only one of the GREDOS objects (1ES0241+622) has got a column density $<$10$^{22}$~cm$^{-2}$.
The algorithm to determine this ``optimal energy band''
will be explained in Sect.~2.1.
Moreover, we have retained only sources with
a BAT flux $\ge 4 \times 10^{-11}$~erg~cm$^{-2}$~s$^{-1}$. This value corresponds to the
count rate threshold used by de la Calle P\'erez et al. (2010) to define
a flux-limited sub-sample (FERO {\it Finding Extreme
Relativistic Objects}) of the RXTE All Sky Survey (\cite{revnivtsev04}).

The aforementioned
upper threshold on $N_H$ ($6 \times 10^{23}$~cm$^{-2}$)
has been chosen to ensure that the obscuring column density
does not prevent the view of the primary emission at the energies of the iron emission
features in any of the GREDOS objects.

The results on the GREDOS sample in this paper will be compared to and jointly discussed
with a sample extracted from the
flux-limited FERO
sample (\cite{delacalle10}). The latter
comprises all the flux-limited FERO AGN (21 sources)\footnote{Ark~120, Ark~564, ESO~198-G24, Fairall~9, H0557-385,
IC~4329A, MCG-2-58-22, MCG-6-30-15, MCG+8-11-11, MR~2251-178,
Mrk~110, Mrk~279, Mrk~509, Mrk~766, NGC~3516, NGC~3783, NGC~4051, NGC~5548, NGC~7314,
NGC~7469, UGC~3973}.
whose X-ray spectrum is obscured by a ``cold'' photoelectric column density
$<$5$\times 10^{21}$~cm$^{-2}$.
The FERO sub-sample so defined is therefore complementary to GREDOS.
The FERO and GREDOS samples together
can be used to test the invariance of the relativistically broadened features with
optically defined spectral type which is predicted by the Seyfert unification scenarios.

The GREDOS sample comprises 13 objects
(Tab.~\ref{tab1}).\footnote{Two objects in GREDOS (MCG-5-23-16, NGC~526A) belong to the FERO flux-limited
sample after de la Calle P\'erez et al. (2010)
due to the slightly different column density thresholds employed by them to define an object as
``X-ray unobscured''.}
\begin{table*}
\caption{The GREDOS sample.}
\label{tab1}
\begin{tabular}{lcccccccc} \hline \hline
Obs.\#     & NED Name       & Type$^a$ & $z$   & $f_{BAT}$$^b$ & $E_{soft}$$^c$ &$r_e$$^d$ & $C_t$$^e$ & T$_{exp}$$^f$ \\ \hline
0006220201 & NGC4507                   & Sy2 & 0.012 & 146.0 &  3.0 &  60/ 67 & 0.35/ 0.5 &  42.0/ 33.2 \\
0067540201 & Mrk348                    & Sy2 & 0.015 & 112.0 &  2.5 &  60/ 40 &   1/  2 &  26.4/ 22.5 \\
0110930701 & NGC4388                   & Sy2 & 0.008 & 203.0 &  3.0 &  68/ 32 &  0.5/  1 &  11.5/  7.3 \\
0144230101 & Mrk6                      & Sy1.5 & 0.019 &  43.7 &  1.0 &  85/ 37 & 0.35/ 0.5 &  36.0/ 31.3 \\
0145670101 & NGC2110                   & Sy2 & 0.008 & 215.0 &  1.5 &  81/ 35 & 0.35/ 0.5 &  34.0/ 34.8 \\
0152940101 & NGC5252                   & Sy2 & 0.022 &  60.1 &  2.0 &  67/ 38 & 0.35/0.35 &  41.0/ 33.3 \\
0202860101 & NGC7172                   & Sy2 & 0.009 & 119.0 &  1.5 &  77/ 38 &   1/  2 &  44.6/ 40.0 \\
0550450301 & 1ES0241+622               & Sy1.2 & 0.045 &  51.0 &  1.0 &  70/ 40 &   1/ 35 &  15.2/ 12.1 \\
0550451501 & GRS1734-292               & Sy1 & 0.021 & 103.0 &  1.0 & 112/ 38 &   1/ 0.5 &  17.2/ 13.5 \\
Tab.~\ref{tab2} & NGC5506                   & Sy2 & 0.006 & 196.0 &  2.0 &  50/ 50 & 0.35/0.35 &  61.0/117.8 \\
Tab.~\ref{tab2} & NGC4151                   & Sy1.5 & 0.003 & 376.0 &  2.2 &  38/ 39 & 0.35/ 0.5 &  87.5/ 71.9 \\
Tab.~\ref{tab2} & NGC526A                   & Sy1.9 & 0.019 &  42.6 &  1.2 &  89/ 38 & 0.35/ 0.5 &  53.7/ 45.4 \\
Tab.~\ref{tab2} & MCG-05-23-016             & Sy2 & 0.008 & 166.0 &  1.8 &  40/ 47 & 0.35/ 0.5 & 118.4/104.8 \\
\hline \hline
\end{tabular}

\noindent
$^a$optical Seyfert type according to the V\'eron-Cetty \& V\'eron catalogue (\cite{veron10})

\noindent
$^b$14--195~keV BAT flux (in units of 10$^{-12}$~erg~cm$^{-2}$~s$^{-1}$) after Cusumano et al. (2010)

\noindent
$^c$low boundary of the energy interval where the spectral fit was performed

\noindent
$^d$size (in arc-seconds) of the source spectrum extraction region (MOS/pn)

\noindent
$^e$threshold (in counts per second) on the single-events, high-energy light curve to
reject intervals of high particle background (MOS/pn)

\noindent
$^f$net exposure time in ks (MOS/pn) after data screening

\end{table*}
It is important to stress that GREDOS is neither a complete nor an unbiased sub-sample of PBC.
XMM-Newton observed some of the GREDOS objects multiple times. We have at first separately analysed the individual
time-averaged spectra with simple photoelectrically absorbed power-law models.
Spectra corresponding to best-fit parameter values consistent within the statistical uncertainties were
merged together. For each source we retained only
the spectrum with the longest exposure time after this merging process.
This is the same procedure followed with the FERO sources (\cite{delacalle10}).
In Tab.~\ref{tab2} we show the list of XMM-Newton Observations IDs
\begin{table}
\caption{List of merged observations for GREDOS sources.}
\label{tab2}
\begin{tabular}{ll} \hline \hline
Source & Merged Obs.\# \\ \hline
NGC~4151 & 0112310101 0112830201, 0112830501 \\
NGC~526A & 0150940101 0109130201 \\
NGC~5506 & 0554170101, 0554170201 \\
MCG-5-23-16 & 0112830301, 0302850201 \\
\hline \hline
\end{tabular}
\end{table}
which were merged together.

\subsection{Data reduction and analysis}

In this paper we discuss data taken with the EPIC cameras only: EPIC-pn (\cite{struder01}), and EPIC/MOS
(\cite{turner01}). Data were reduced using SASv10 (\cite{gabriel03}), using the calibration files available in
August 2010. Calibrated event lists were generated using the reduction meta-tasks {\tt e[mp]proc}. Source
spectra were extracted from circular extraction regions centred around the source X-ray centroid. The size
of the source scientific products
extraction region $r_e$, as well as the count rate thresholds $C_t$ employed
to reject intervals of high particle background
were determined through an iterative process to maximise the net source spectra
signal-to-noise ratio (see
Tab.~\ref{tab1}). To measure the particle-induced background we used the whole field-of-view, high-energy
(MOS: $E >$10~keV;
pn: 10~keV~$\le E \le$~12~keV) light
curve binned to $\Delta t$=10~s. Background spectra were extracted from off-axis circular regions on the
same CCD as the source, and additionally at the same {\tt RAWY} position in the EPIC-pn camera to ensure that
the same charge transfer efficiency correction applies, as this correction depends on the distance from the readout node.
Response files for each spectrum were generated using the SAS tasks {\tt arfgen} and {\tt rmfgen}.

Spectra were rebinned in order to ensure that: a) each background-subtracted spectral channel contains at least
50 counts; b) the instrumental intrinsic resolution is oversampled by a factor not larger
than 3. These conditions ensure that we can use the $\chi^2$ as a goodness-of-fit test in the forthcoming spectral
analysis. Spectral fits were performed in the energy range [$E_{soft}$,10.0~keV]. $E_{soft}$
(see Tab.~\ref{tab1}) was determined
for each spectrum after visual inspection as the energy where excess emission above the extrapolation
of the photoelectrically absorbed power-law emission starts to dominate. The rationale behind this choice is
driven by the empirical evidence that this soft excess is associated to diffuse emission extended on
scales as large as a few kilo-parsecs, most likely due to gas associated to the Narrow and Extended Narrow
Line Regions (\cite{bianchi06}), with contribution by strong starburst emission in a few objects
(\cite{guainazzi09}). The modelling of this component is therefore fully decoupled from
that of the the primary nuclear emission.

In this paper we quote statistical errors at the 90\% confidence level for one interesting parameter unless otherwise
specified. In order to calculate luminosities we use the following cosmological parameters:
$H_0$=70~km~s$^{-1}$~Mpc$^{-1}$, $\Lambda_0$=0.73, $\Omega_M$=0.27 (\cite{bennett03}).

\subsection{Spectral model definition}

The main feature characterising X-ray spectra of obscured AGN is a soft X-ray photoelectric cut-off
due to high column densities of obscuring gas (\cite{awaki91}). We therefore
first fit the GREDOS spectra with a simple baseline continuum constituted by a power-law modified by
photoelectric absorption trough cold matter. However,
moderate-resolution spectroscopy of nearby Seyfert galaxies unveils further spectral complexity,
associated to reprocessing of the primary nuclear continuum (\cite{pounds90,nandra94,turner97}). This evidence
led us to add to the model continuum a Compton-reflection component (model {\tt pexrav} in {\sc Xspec};
\cite{magdziarz95}) as well as a number of Gaussian emission line profiles to account for K$_{\alpha}$
(with its Compton shoulder; \cite{matt02}) and K$_{\beta}$
fluorescent lines from neutral iron (\cite{turner97}), and for recombination lines from He- and H-like
iron (\cite{netzer98,bianchi02}). This ``non-relativistic'' baseline model can be described as follows:
$$
M(E) = e^{-\sigma_p(E) N_H} \times [ A_1 E^{-\Gamma} + \Sigma_i N_i G_i(E_i) +
$$
$$
A_2 C_S(E) ] + A_3 C_R(E,\Gamma) 
$$
where $\sigma_p$ is the photoelectric cross-section, $G_i$ are Gaussian profiles, $C_S$ is the Compton
shoulder, $C_R$ is the
Compton reflection continuum, and $A_i$ and $N_i$ are normalisation constants. We used the {\tt ztbabs}
implementation for the photoelectric absorption in {\sc Xspec}.
In the baseline, as well as in all subsequent more complex models, we made the following
assumptions:

\begin{itemize}

\item the photon index $\Gamma$ of the power-law and of the Compton reflection continuum has been constrained to be
the same

\item the power-law is modified by a
high-energy cut-off beyond the energy bandpass of the EPIC cameras. It
have been held fixed to 150~keV (\cite{risaliti02})

\item the inclination of the Compton reflection component with respect to the line-of-sight
has been held fixed to 45$^{\circ}$. This parameter is
degenerate with the relative normalisation between the Compton reflection and the primary continuum, therefore
this assumption does not unnecessarily constraint the parameter space of other spectral parameters

\item we have assumed solar abundances according to Anders \& Grevesse (1989)

\item the centroid energy of the Gaussian profiles has been held fixed to values as dictated by the atomic
physics: 6.4~keV, 6.966~keV, 7.058~keV for {\sc Fe i} K$_{\alpha}$, {\sc Fe xxvi}, and
{\sc Fe i} K$_{\beta}$, respectively. The intensity of the {\sc Fe i}  K$_{\beta}$ has been constrained not to
exceed 16\% that of the K$_{\alpha}$. For the {\sc Fe xxv} transitions we have allowed the
centroid energy to vary between 6.6364 and 6.7000~keV, corresponding to the forbidden and the
resonant transitions, respectively

\item the intrinsic width of the Gaussian profiles has been assumed to be the same for a given object, and
consistent with the instrumental resolution

\item following Matt (2002), the Compton shoulder was modelled with a rectangular box of width 50~eV and
intensity constrained not to exceed 20\% of the intensity of the corresponding K$_{\alpha}$ profile

\end{itemize}

Only after these additional components were included (and kept, if their inclusion improved the quality
of the fit at a confidence level $>$90\%; \cite{lampton76}), we tried to fit any further spectral complexity with
relativistically blurred disc components. For this scope, we have used models extracted from the {\tt ky} suite
(\cite{dovciak04}), based on a common ray-tracing subroutine aiming at describing the X-ray emission of black-hole
accretion discs in the strong gravity regime. As any emission line should come accompanied by a relativistically
blurred disc continuum, we have used a combination of the models {\tt kyrline} and {\tt pexrav} where the latter
was convolved with a relativistic kernel {\tt kyconv}. In both {\tt kyrline} and {\tt kyconv} the blurring effect
depends on the following parameters:
a) the black hole spin $a$ (comprised between 0 and 0.9982 in dimensionless relativistic units);
b) the inner ($r_{in}$) and outer ($r_{out}$) radius of an annular region on the disc where
the line photons are emitted; c) the index $\alpha$ of the
radial dependence of the emissivity per unit area, $\kappa$, in the local frame
comoving with the disc: $\kappa (r) \propto r^{-\alpha}$; d) the
``disc inclination'' $\imath$, i.e. the angle between the normal to the disc plane and
the line-of-sight; e) the rest frame
energy (assumed monochromatic) of the photons emitted by the disc, $E_c$.
Although the line normalisation in {\tt kyrline} is expressed in units of
photons~cm$^{-2}$~s$^{-1}$ integrated over the whole profile, we will also use the line
Equivalent Width (EW), i.e. the line intensity normalised to the underlying continuum at
6.4~keV, to ease comparison against theoretical predictions. 
Following the approach described in Guainazzi et al. (2010), we fixed the values of $r_{in}$ and $r_{out}$ to
the innermost circular stable orbit (e.g. $r_{in} = 1.24$$r_g$ for a maximally spinning black hole) and 
$r_{out}$=400$r_g$, respectively. Constraints on the black hole spin could not be derived for any of
the GREDOS sources. We therefore assumed a maximally spinning black hole in all the subsequent fits.

For objects where
this model did not provide an adequate fit we changed the obscuration law. We have tried
the following additional prescriptions for the X-ray obscuration: a) two ``cold'' absorbing systems, one covering
the primary and the relativistically broadened reprocessing components, and the other covering the whole
model; b) a partial covering absorber covering solely the primary and the
relativistically broadened reprocessing components; c) one ionised absorber, calculated according to a grid
of {\sc Cloudy} models (\cite{ferland98}) as described in Bianchi et al. (2010); d) the combination of
a ``cold'' and a ``ionised'' absorber.

For each source we have selected the model which yields the best reduced $\chi^2$.
We consider in the following
a relativistic line as detected if the variation of $\chi^2$ associated with the inclusion of the
{\tt kyrline} component in the spectral model is significant at the 5-$\sigma$ level, to be consistent with
the criterion adopted by \cite{delacalle10}.

\subsection{Spectral results}

\begin{figure}
\hspace{-0.75cm}
\includegraphics[height=100mm,angle=90]{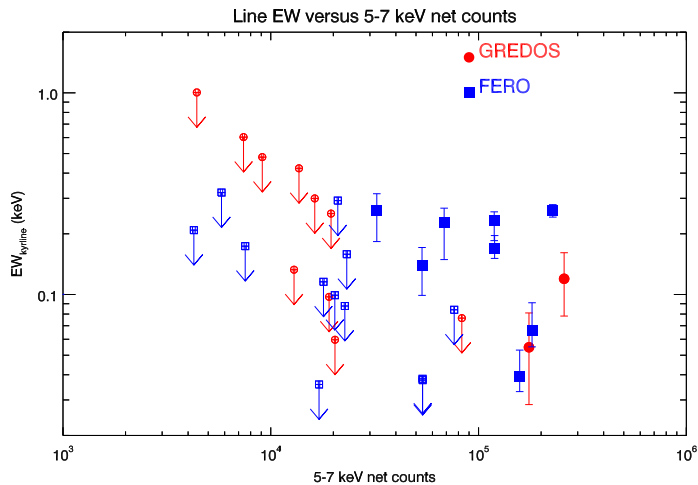}
\caption{
Relativistic line EW as a function of the background-subtracted 5--7~keV counts
in the GREDOS ({\it circles}) and FERO ({\it squared})
samples sources. {\it Empty symbols} indicate upper limits. The {\it lines} represent
the sample sensitivity for three different values of inclination angle, assuming
$\alpha = 2.5$: {\it dashed line}: 10$^{\circ}$; {\it dot-dashed line}: 40$^{\circ}$;
{\it dot-dot-dashed line}: 80$^{\circ}$.
}
\label{fig6}
\end{figure}

Tab.~\ref{tab3}, and Tab.~\ref{tab4} of Appendix~A
summarise the results of our spectral fits on the GREDOS sample.
At the adopted 5-$\sigma$ confidence level, a broadened relativistic line is detected in only two
sources. In both cases this detection was reported elsewhere: MCG-5-23-16 (\cite{dewangan03,reeves07,delacalle10}),
and NGC~5506 (\cite{guainazzi10}).
Our non-detections are also consistent with published reports: Mkn~6
(\cite{schurch06}), NGC~4151 (\cite{schurch03}), NGC~4388 (\cite{beckmann04}), NGC~5252 (\cite{dadina10}).
The detection fraction in GREDOS ($\simeq$15\%) is nominally lower then in samples of unobscured AGN.
However, the GREDOS sample is incomplete. This prevents a statistically robust comparison of this number with
less unbiased samples such as FERO. Moreover, most of the GREDOS sources are underexposed.
In Fig.~\ref{fig6} the EW of the relativistic line is plotted against the background-subtracted counts
in the 5--7~keV energy band. As already observed by other authors (\cite{guainazzi06,delacalle10}),
the detection is a strong function of the number of net source counts in the hard X-ray band.
All broad line detections in the FERO and in the GREDOS samples correspond to objects with more then 30,000
net source counts in the 5--7~keV energy band. Only three sources in GREDOS satisfy this
empirical criterion: MCG-5-23-16, NGC5506 (the two detections), and NGC~4151. Intriguingly enough,
above 100,000 net counts one measures a 100\% detection fraction.

\begin{figure*}
\hbox{
\includegraphics[height=57mm,angle=-90]{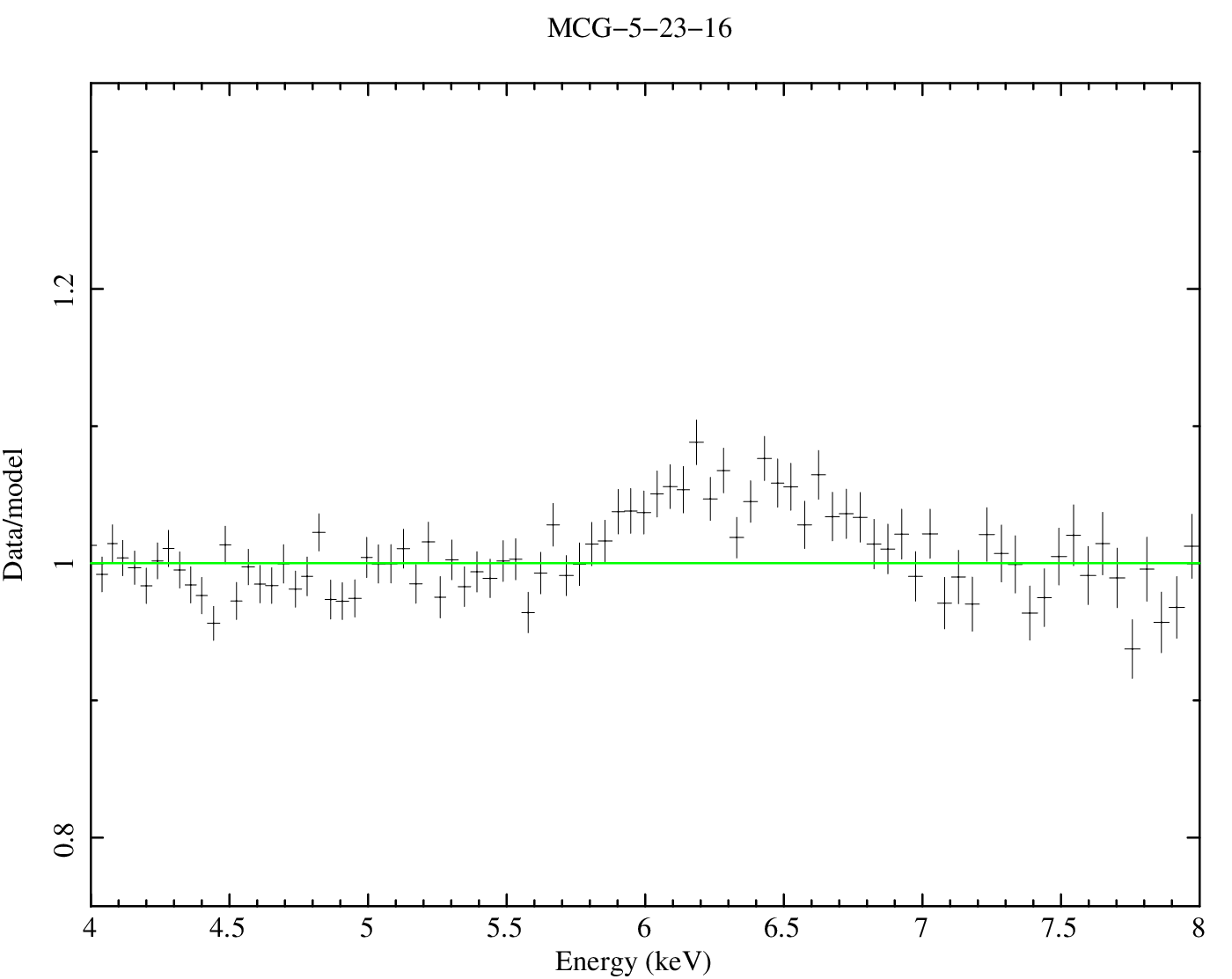}
\includegraphics[height=57mm,angle=-90]{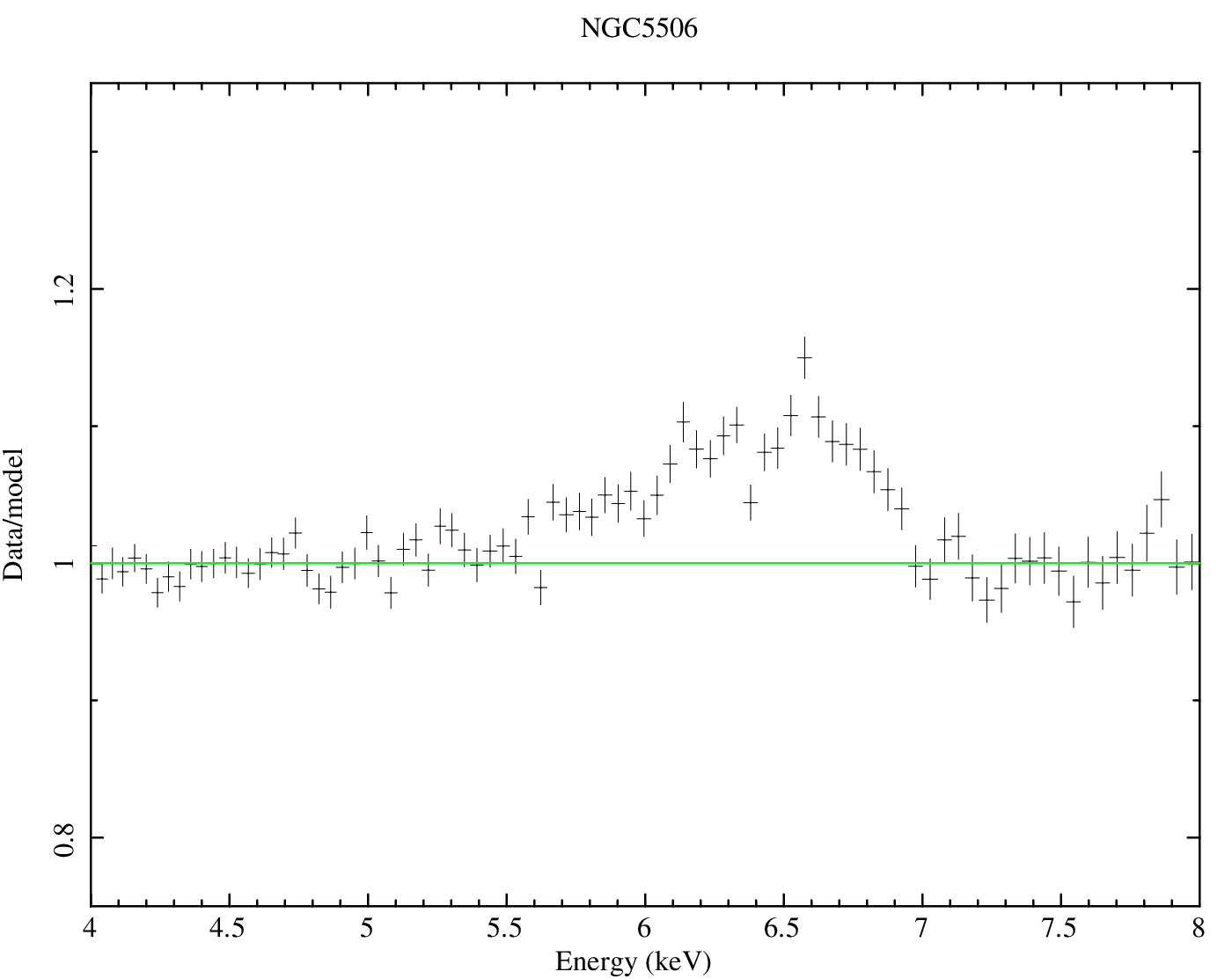}
\includegraphics[height=57mm,angle=-90]{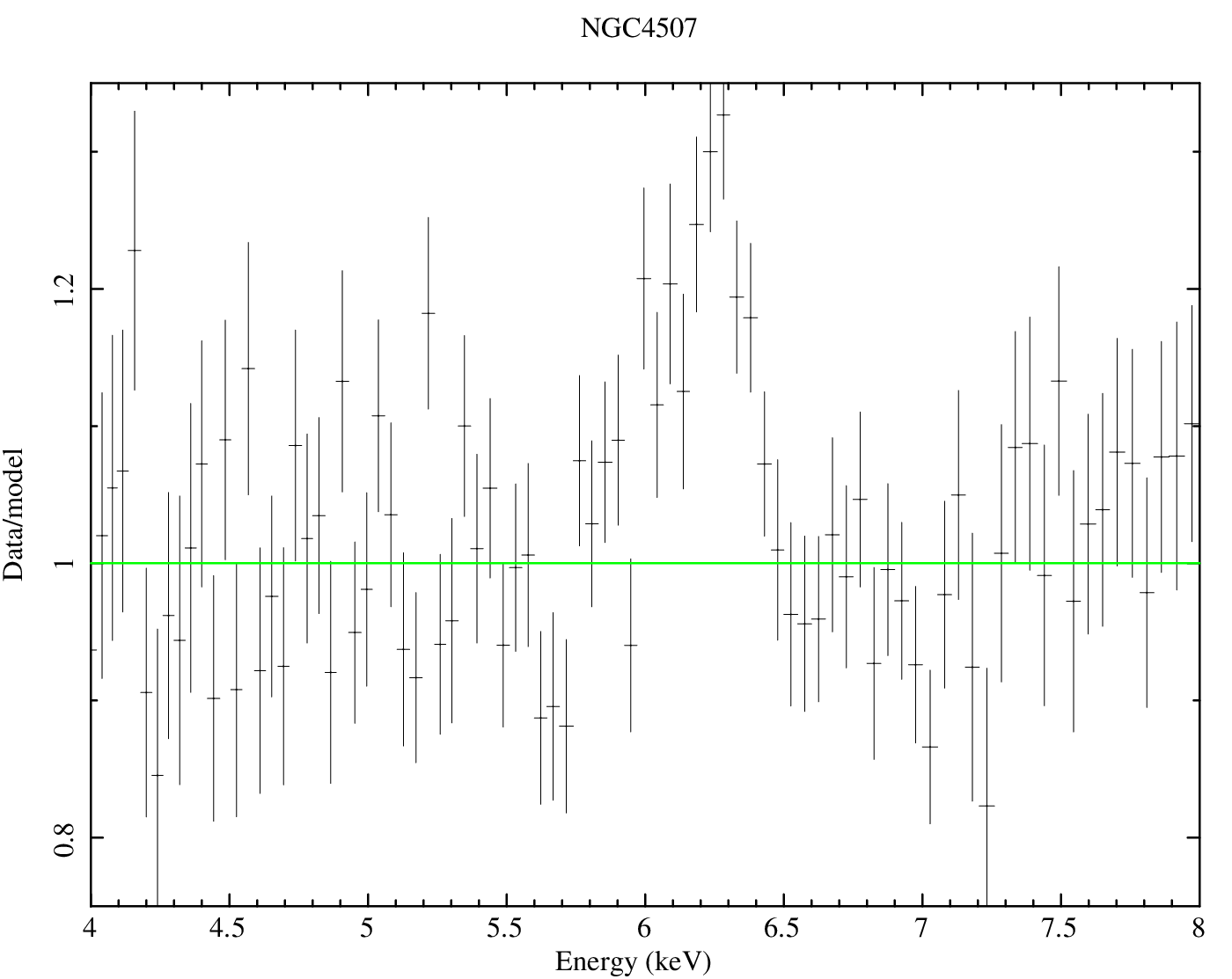}
}
\caption{
Profiles of the relativistically broadened iron lines detected in the GREDOS sample.
They represent data/model ratios once the {\tt kyrline} component
is removed from the best-fit model. The lines in MCG-5-23-16 and NGC~5506 are detected
at a confidence level $\ge$5$\sigma$ in the relativistically blurred
line scenario. The line in NGC~4507 is detected at a confidence
level $\simeq$3$\sigma$ only.
}
\label{fig3}
\end{figure*}

\begin{figure}
\hspace{-0.75cm}
\includegraphics[height=100mm,angle=90]{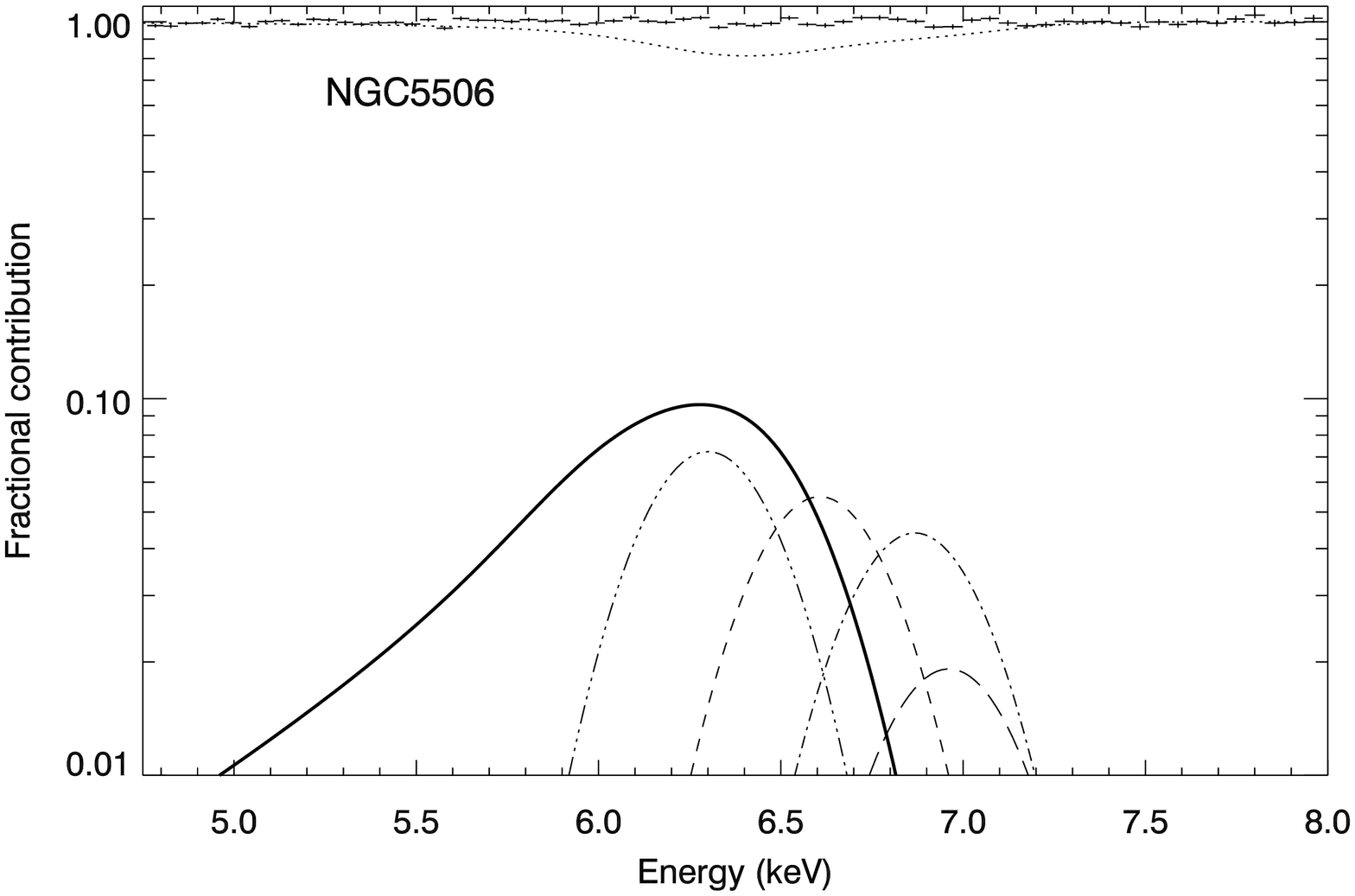}

\hspace{-0.75cm}
\includegraphics[height=100mm,angle=90]{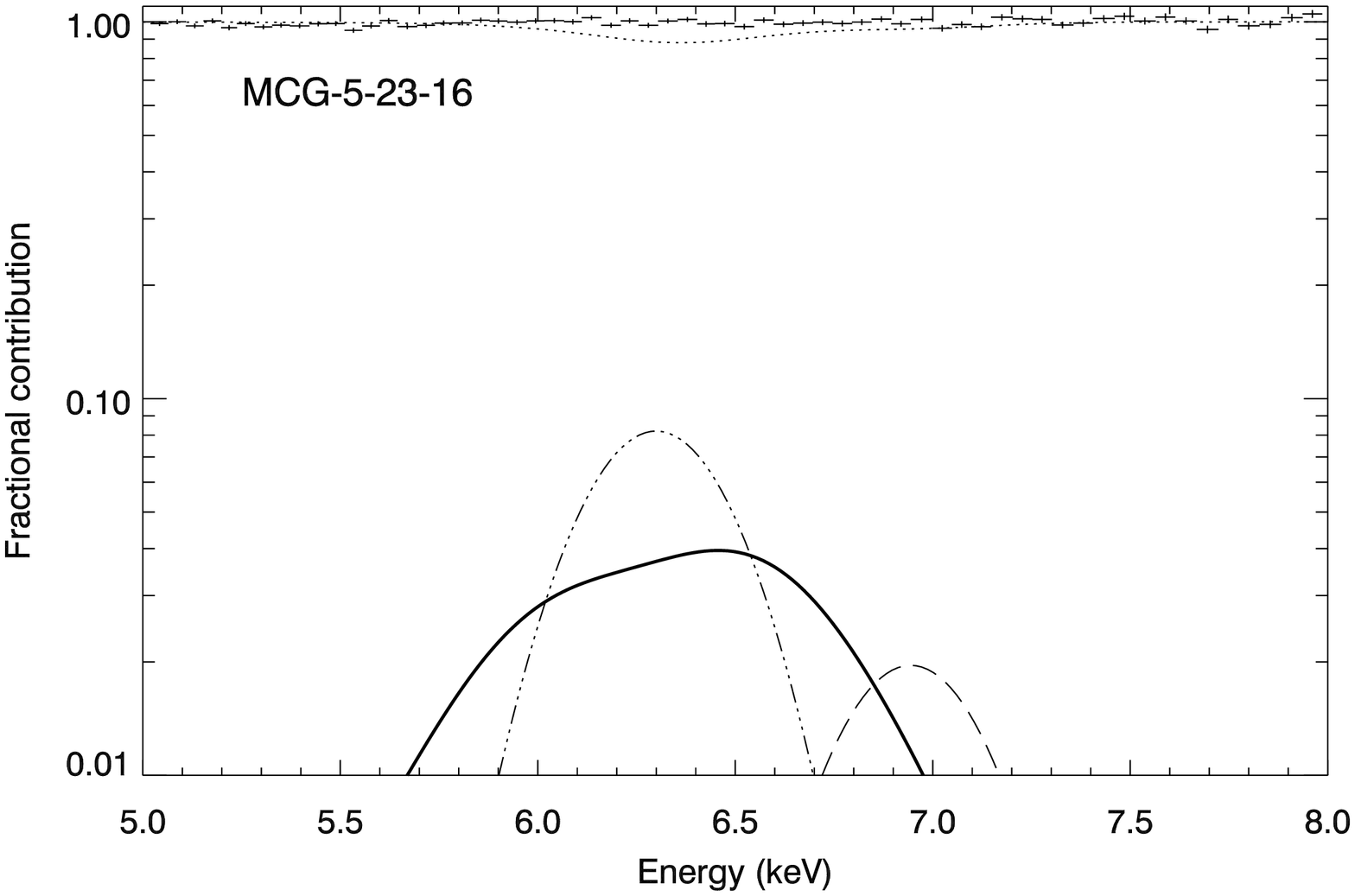}
\caption{
Fractional contribution of the best-fit model components to the total
best-fit model. {\it
Dotted line}: continuum; {\it solid line}: relativistically broadened Fe;
{\it dashed-multi-dotted line}: unresolved Fe{\sc i} K$_{\alpha}$;
{\it dashed line}: unresolved Fe{\sc xxv} recombination;
{\it dashed-dotted line}: unresolved Fe{\sc xxvi} recombination;
{\it dashed line}: unresolved Fe{\sc i} K$_{\beta}$ line. All components
have been smoothed with a Gaussian kernel ($\sigma_{kernel}$=150~eV)
to reproduce the effect of the EPIC-pn intrinsic energy resolution. The
{\it crosses} represents the data/model ratio against the best-fit model.
}
\label{fig1}
\end{figure}

At a lower confidence level (3$\sigma$) an additional broad line is detected in a GREDOS source:
in NGC~4507, by contrast to \cite{matt04}. Its parameters are: $\alpha \le 1.7$, $\imath = 10\pm^{4\circ}_8$,
$EW = 128\pm^4_{60}$~eV  if $E_c \equiv 6.4$~keV. We will not include this measurement in
the joint analysis of the FERO and GREDOS detections (see Sect.~2.4).
The profiles of the iron lines in GREDOS are shown
in Fig.~\ref{fig3}.
These line profiles are obtained from the data/model ratio against the best-fit
once the relativistic line profile ({\tt kyrline} component) is removed.
In Fig.~\ref{fig1} we show the fractional contribution
of the continuum as well as of the narrow-band emission features to the total
best-fit model in the two GREDOS objects, where a broad line is detected at a confidence level
$\ge 5$$\sigma$.

\subsection{FERO+GREDOS detections}

Hereafter we will consider the detections in the FERO and GREDOS samples
jointly (cf. Sect.2 for the definition of the FERO sample in this paper).
In this sample, 12 broad iron lines are detected at confidence level
$\ge$5$\sigma$ (cf. Tab.~2 in \cite{delacalle10}, and Tab.~\ref{tab4} in this paper).

\begin{figure}
\hspace{-0.75cm}
\includegraphics[height=100mm,angle=90]{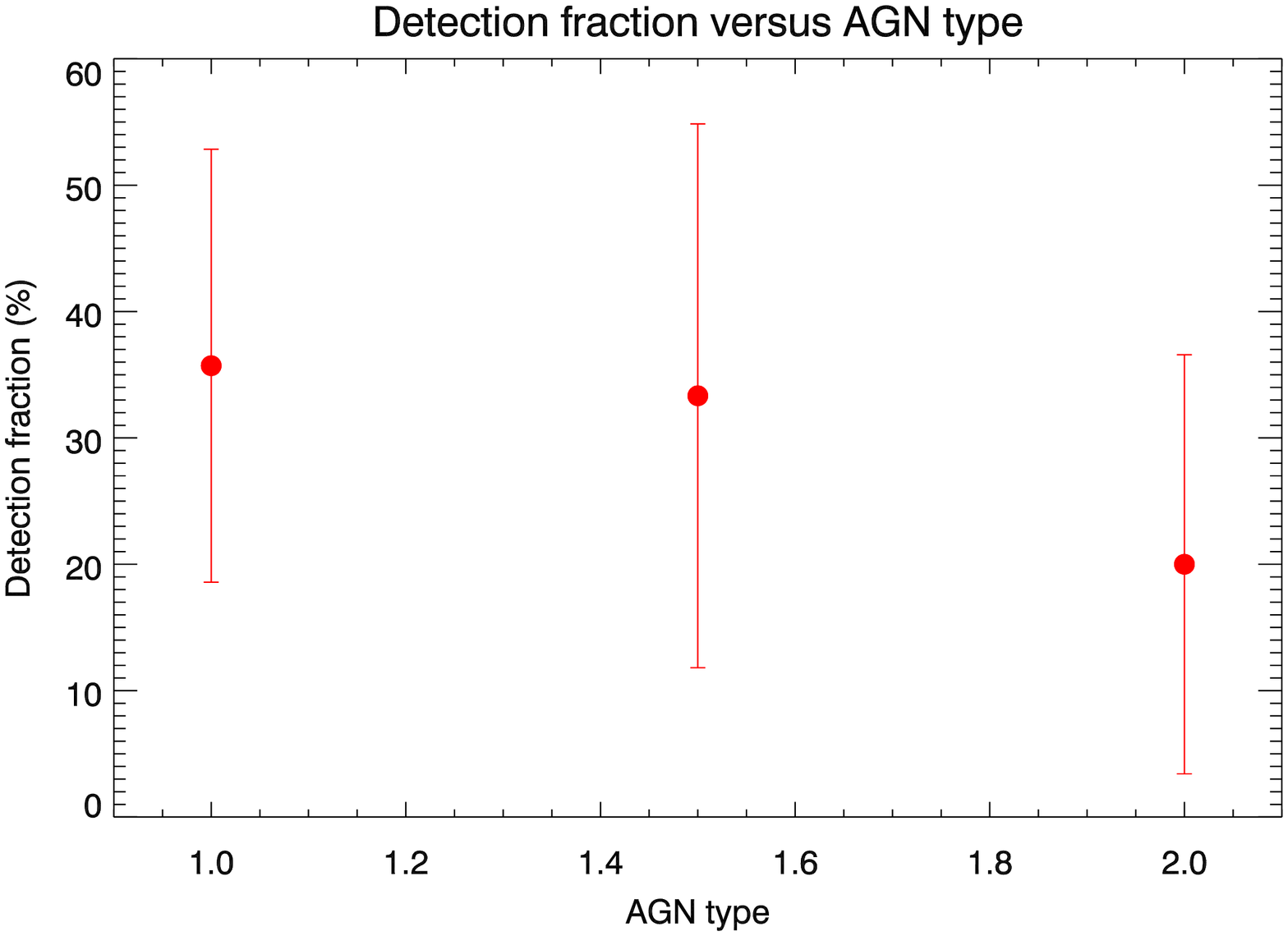}

\hspace{-0.75cm}
\includegraphics[height=100mm,angle=90]{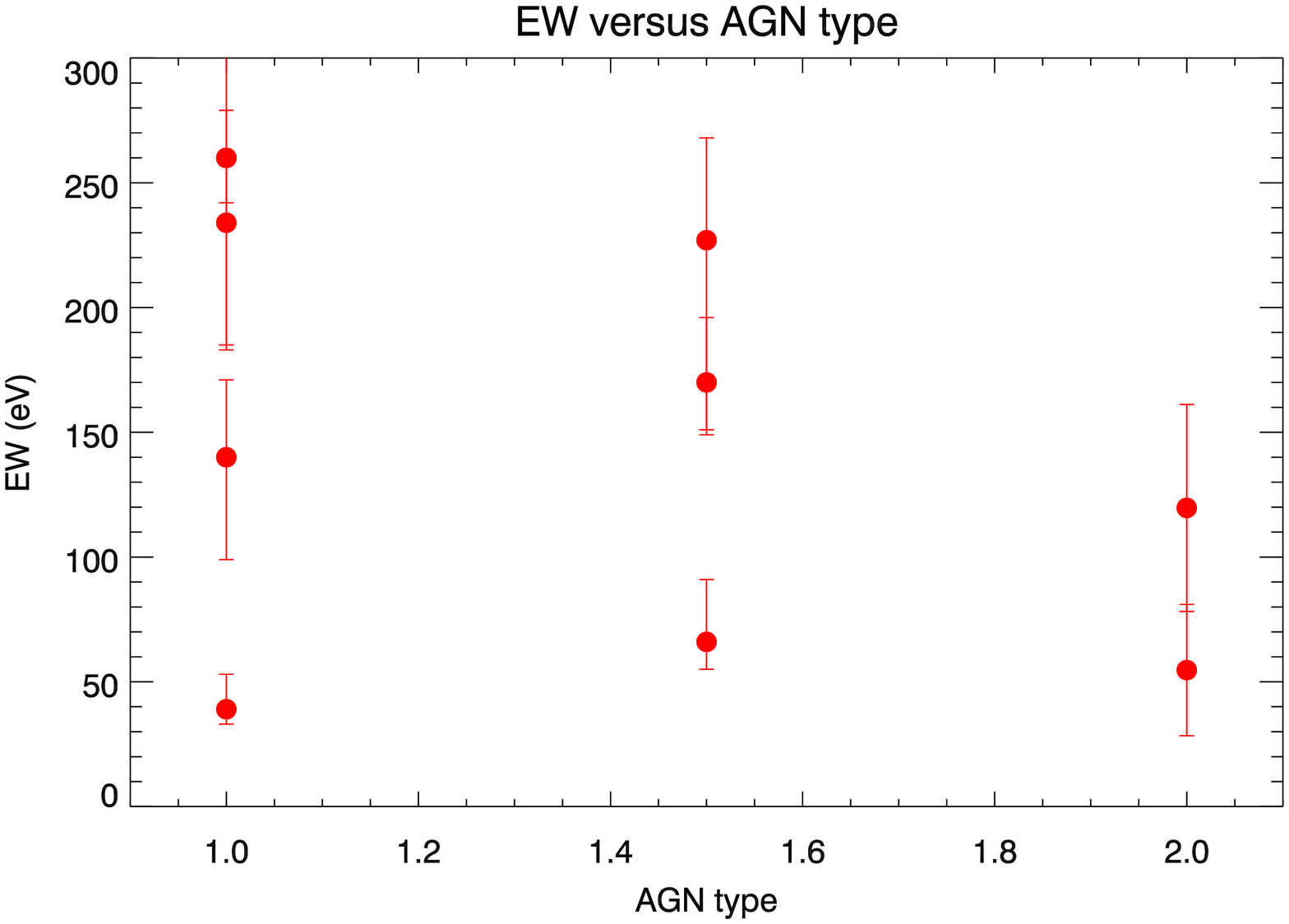}
\caption{
Relativistic line detection fraction ({\it upper panel}) and EW ({\it lower panel}) as a function of
optical type for the sources of
joint FERO+GREDOS samples.
}
\label{fig7}
\end{figure}

Once the FERO and GREDOS detections are considered jointly, neither the detection fraction, nor
the line EW are dependent on
the optical type (Fig.~\ref{fig7}), despite a nominal trend for broad lines to
be more common in ``type 1'' Seyferts.
In this paper we use the optical type as defined in the Veron-Cetty \& Veron (2010) catalogue.
We use the optical classification to situate, at the best of our knowledge, the FERO+GREDOS sources
in the framework of the standard Seyfert unification scenario as reviewed in, {\it e.g.}, Antonucci (1993).
For this purpose, objects which exhibit broad polarised Balmer lines and/or
weak IR broad lines, but whose optical spectrum is typical of obscured AGN
are still considered ``type 2''\footnote{MCG-5-23-16, Mkn~348, NGC~2110, NGC~4388, NGC~4507,
NGC~5506, NGC~7314}. For sake of simplicity
we include Veron type 1.9 objects into the ``type 2'' class, Veron type 1.2 objects and Narrow Line Seyfert
galaxies\footnote{Mkn~110, Mkn~766, NGC~4051} into the ``type 1'' class, hereafter.

Following the same approach as in, {\it e.g.}, Guainazzi et al. (2010) we use the $\alpha$ parameter in
{\tt kyrline} as a measurement of the importance of relativistic effects in shaping the profile of the
iron K$_{\alpha}$ emission line. For the same value of the inner and outer radius of the line-emitting region being
equal, higher $\alpha$ values indicate that line photons
are emitted in regions closer to the accretion disc innermost stable orbit. For $\alpha \le 2$ ({\it i.e.}
when a large fraction of the accretion disc is responsible for the line production) the choice of
the line-emitting region outer radius (which, we remind, is fixed to 400$r_g$ in our spectral models)
may affect the best-fit values of other line parameters. This effect is discussed
by Guainazzi et al. (2010) for the case of NGC~5506.

\begin{figure}
\hspace{-0.75cm}
\includegraphics[height=100mm,angle=90]{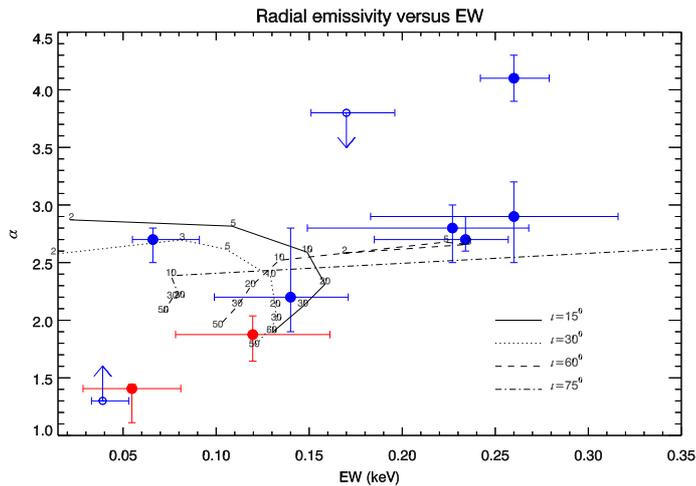}
\caption{
$\alpha$ versus EW for the iron K$_{\alpha}$ relativistic lines detected in the FERO and GREDOS
samples. {\it Empty symbols} indicate upper limits. The {\it lines} represent the predictions
of a lamppost model for different values of the X-ray source height ({\it small numbers along the
curves} in units of gravitational radii) and disc inclinations ({\it line styles}; cf.
the inset label).
}
\label{fig8}
\end{figure}

In the joint FERO and GREDOS detections one observes an apparent correlation between
$\alpha$ and EW (Fig.~\ref{fig8}).
A fit to the data using an extension of the regression method on censored data
originally described by Schmitt (1985) and Isobe et al. (1986) yields:
$\alpha = (0.44 \pm 0.19) + (2.0 \pm 0.8) \times \log(EW)$, when $EW$ is expressed in keV.
The Spearmann rank coefficient on the same data
yields a correlation probability of $99.7\pm^{0.2}_{17.4}$\%,
where the nominal value is calculated using only bracketed measurements, and the error
bars reflect the range of values that censored data could cover on the y-axis (only
subject to the best-fit model parameter
restriction: $\alpha \le 6$). Taking into account the large uncertainties,
the correlation is statistically marginal.
In the same figure, we compare the measurements
with the predictions of an axisymmetric lamppost model
(\cite{dovciak04}), where a point-like source of X-ray radiation is located at variable
heights along the black hole spin axis. We simulated EPIC-pn spectra according to this
model, and fit them using the same baseline model as defined in Sect.~2.2. Moderately
relativistic ($\alpha \approxlt 3$) lines
can be explained in this framework if the accretion disc is seen in different objects
under a wide range of different inclination angles and heights.
Interestingly enough, it is impossible to get $\alpha > 3$ in this scenario.
Recently, examples
of extreme relativistic lines have been published, whose profiles are consistent
with very steep ($\alpha \approxgt 5$) specific emissivity radial dependencies
(\cite{wilms01,fabian09,ponti10}). The main driver for some of these results is the very
smooth shape of the soft excess, implying extreme relativistic blurring of narrow-band
spectral features imprinted by disc reflection on the soft X-ray ($E \approxlt 2$~keV)
spectrum. As pointed out by, {\it e.g.}, \.Zycki et al. (2010) steep profiles
can be obtained only if the X-ray source is not located on the black hole spin
axis, because they require the combination of dynamical Doppler shift and frame dragging
within the black hole ergosphere.

\begin{figure}
\hspace{-0.75cm}
\includegraphics[height=100mm,angle=90]{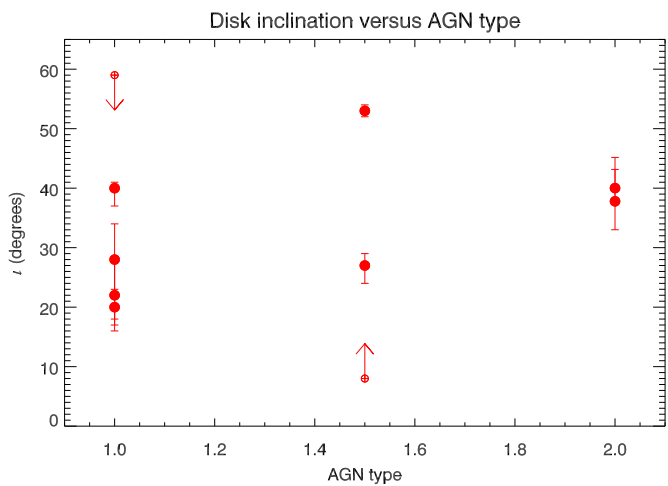}
\caption{
Accretion disc inclination angle $\imath$
as a function of the optical type for the joint GREDOS and FERO control
sample discussed in this paper.
}
\label{fig9}
\end{figure}

There is no significant difference between the distribution of disc inclination angles between the GREDOS and
the FERO sample sources, neither a dependency of the disc inclination angle on the optical spectral type
(Fig.~\ref{fig9})
or host galaxy inclination.
Although the line detections in type~2 Seyfert correspond to $\imath \ge 38^{\circ}$
(MCG-5-23-16, \cite{reeves07}; NGC~5506, \cite{guainazzi10}), high-inclination lines are measured in type~1
objects as well (Mkn~509, MCG-6-30-15; \cite{delacalle10}). Moreover, type~1.5 objects invariably exhibit
inclination angles lower then 40$^{\circ}$, ruling out a simple relation between the inclination of the
nuclear obscuring matter (as measured by the optical type)
and that of the accreting matter. 

Care must be exercised in interpreting direct measurements of the disc inclination
through the profile of a relativistically broadened line, though.
Measurements of the
inclination angles through X-ray spectroscopy in the relativistically blurred line
scenario are still affected by large systematic
uncertainties. Even in the best studied case (the high-quality ``long-look'' EPIC spectrum of MCG-6-30-15
taken in 2002) published measurements of the relativistic line profiles
by different authors yield values of the disc inclination angle
ranging from 20$^{\circ}$ to 48$^{\circ}$ (\cite{brenneman06,fabian02,branduardi01,reynolds04}).

\section{The quest for the ultimate driver of relativistic effect strength}

A number of AGN exhibit relativistically broadened iron lines with an EW $\ge$300~eV (cf. Sect.~1). Such an extreme
strength is difficult to reconcile with theoretical calculations of the EW expected from standard X-ray illuminated
relativistic accretion discs (\cite{george91}), unless the primary emission illuminating the disc is highly
anisotropic. Mechanisms based on strong relativistic effects occurring when the source is located a few
gravitational radii from the black hole event horizon have been proposed (\cite{martocchia02,miniutti04}).
Indeed, light bending coupled with a variable height of the X-ray primary sources on the disc plane
can explain at the same time the extreme strength of relativistic reprocessing spectral components
(\cite{fabian09,ponti10}),
and their lack of
response to the primary continuum variation (\cite{miniutti04}).

A standard plane-parallel X-ray illuminated accretion disc covering a 2$\pi$ solid angle should produce an
EW$\sim$150~eV for a typical AGN X-ray spectral shape and solar abundances (\cite{george91}). However,
many of the FERO+GREDOS sources exhibit EWs significantly lower than this. Are there alternatives to
explaining this large dynamical range in terms of disk solid angle as seen by the primary source?
We will hereby follow a falsification approach to address this problem: we will assume that
for ``standard'' $EW \le 300$~eV AGN the solid angle is on the average the same, and analyse which
consequences this assumption bears on the correlation between EW and other observables.

We first mention the most fundamental limitation of our quest.
High-EW reprocessed emission lines can be produced if the disc photosphere is highly ionised
(\cite{zycki94}). The data currently do not allow us to test this hypothesis.
Fits with a relativistic profile corresponding to being dominated by Fe He-like emission is preferred
in 3 out of 11 FERO (\cite{delacalle10}), in none of the GREDOS detections.
Although in these cases the EW is tendentiously larger then
the distribution average (140-230~eV), this parameter space is not exclusively occupied by them. Moreover,
there is no correlation between the line EW and the X-ray luminosity or the accretion rate normalised to
the Eddington value. This suggests that disc photoionisation probably does not play a decisive role in
determining the strength of relativistic reprocessing for our sample. Much better data quality would be
required, however, to resolve the degeneracy between the bulk ionisation state of the line emitting region
and other parameters.

Once we neglect the disc ionisation, the EW of a relativistically broadened iron line depends primarily on:

\begin{enumerate}

\item the photon index of the illuminating continuum, $\Gamma$

\item the disc inclination, $\imath$

\item the iron abundance, $Z_{Fe}$

\item the ``reflection fraction'', a parameter which for purely isotropic emission is proportional to
the solid angle subtended by the disc at the X-ray source

\end{enumerate}

Approximated formulae for the dependency on the first three parameters are reported by Nandra et al.
(2007)\footnote{Although the formula for the dependency of the EW on the spectral index contains
a typo. The correct formula is 
$EW = EW_0 \times [ 9.66 \times (\Gamma^{-2.80}) - 0.56 ]$ (Nandra, private communication),
where $EW_0$=144~eV.}. In this Section we will assume initially that the reflection fraction is the same
for all objects in the FERO+GREDOS sample and test whether the observed distribution of EW can be
reproduced by any of the other three parameters.

Very few measurements of the iron abundance are available in the literature for the objects of our sample.
Following the approach after Ballantyne (2010) we have therefore decided to calculate $Z_{Fe}$ using the
correlation with the accretion rate in Netzer \& Trakhtenbrot (2007).

\begin{figure*}
\hbox{
\includegraphics[height=80mm,angle=90]{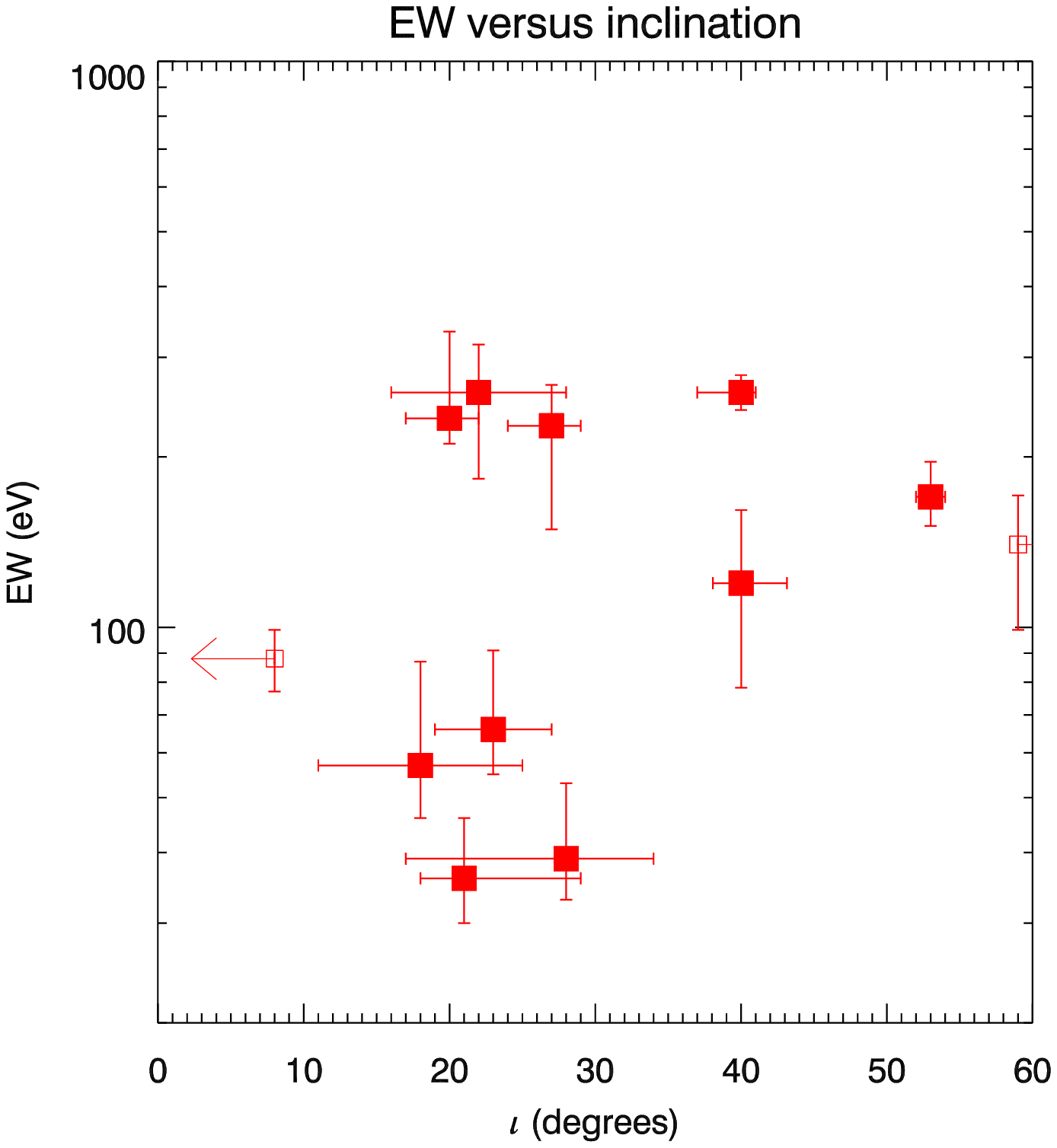}
\hspace{1.0cm}
\includegraphics[height=80mm,angle=90]{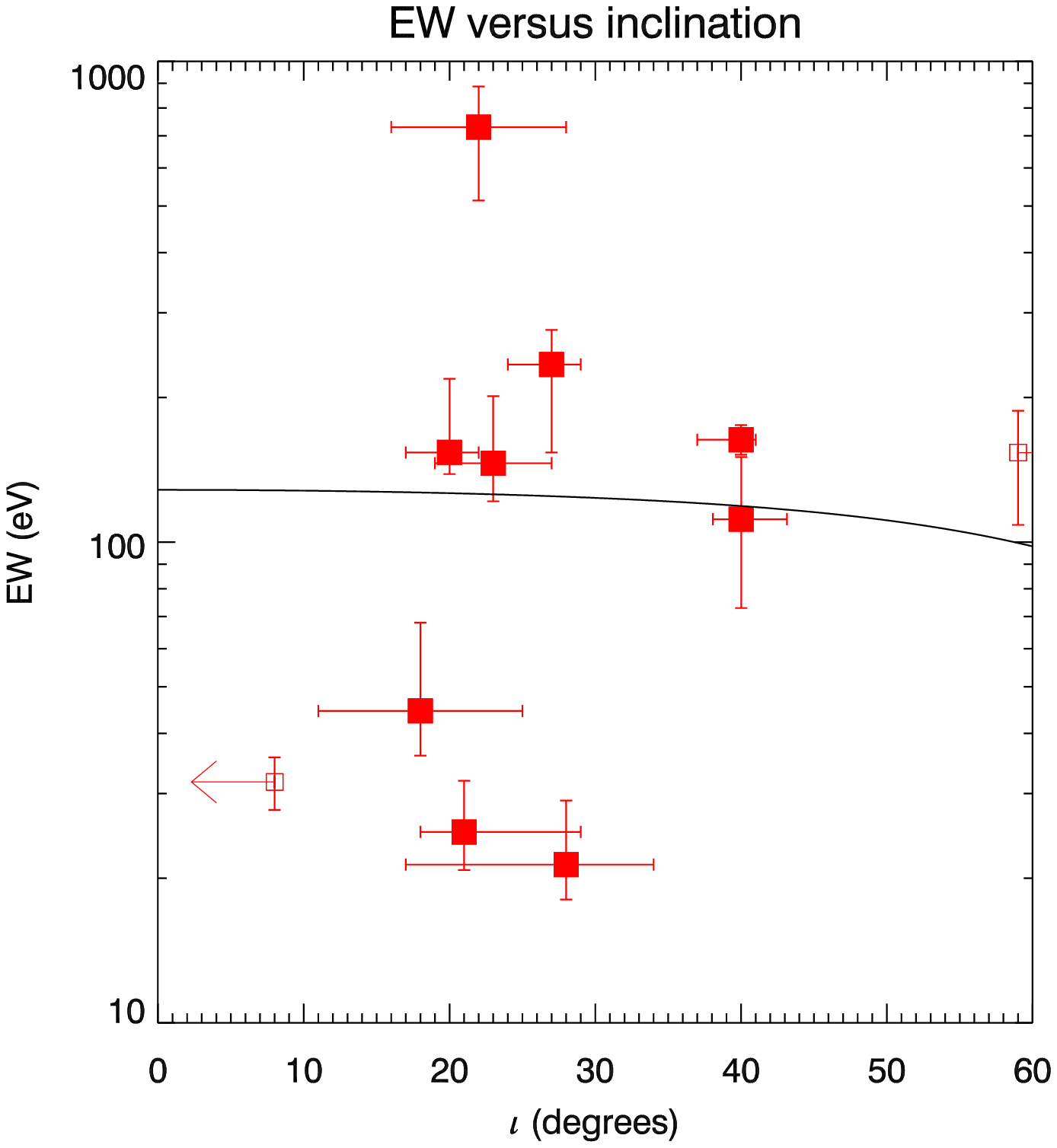}
}
\hbox{
\includegraphics[height=80mm,angle=90]{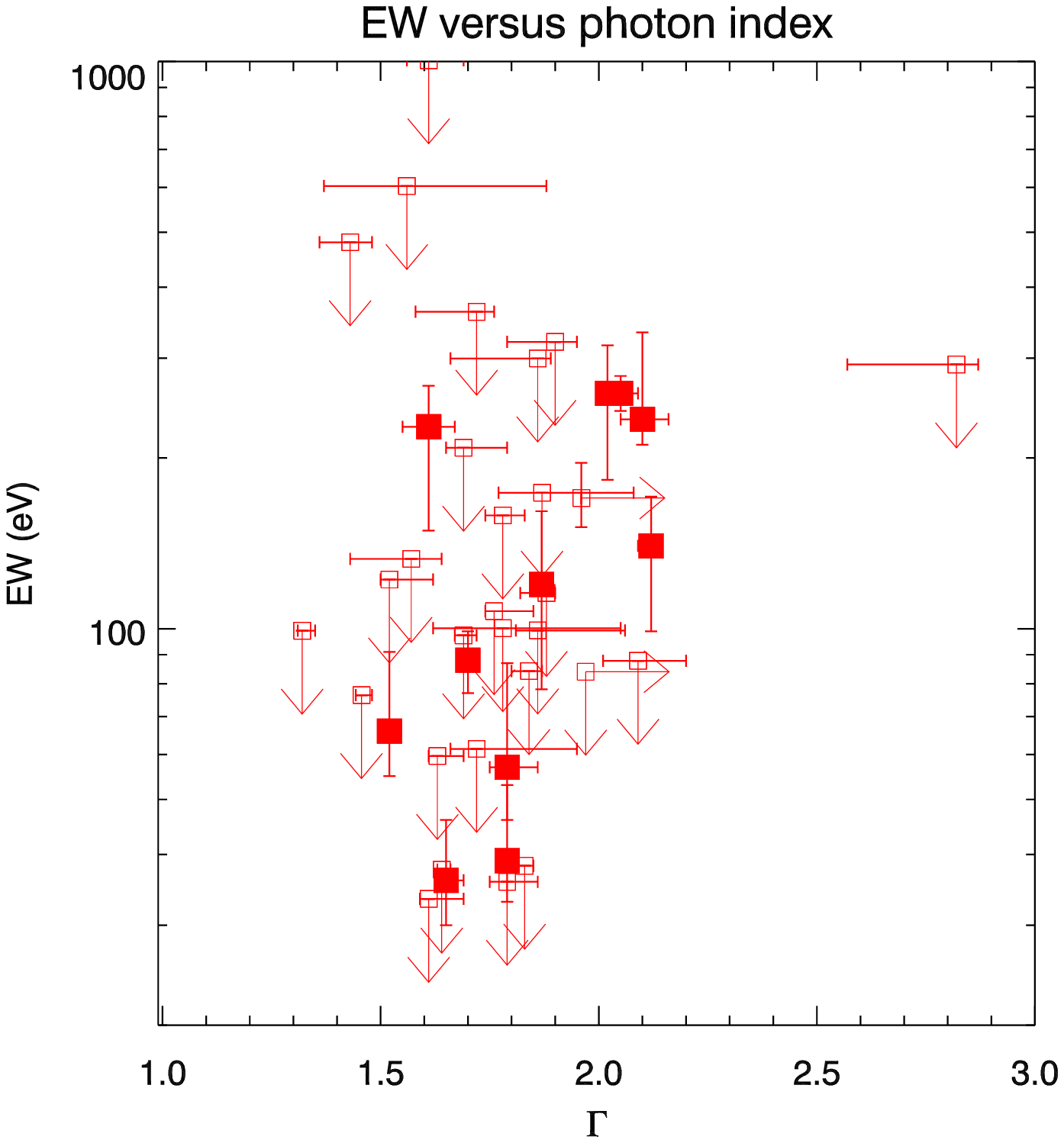}
\hspace{1.0cm}
\includegraphics[height=80mm,angle=90]{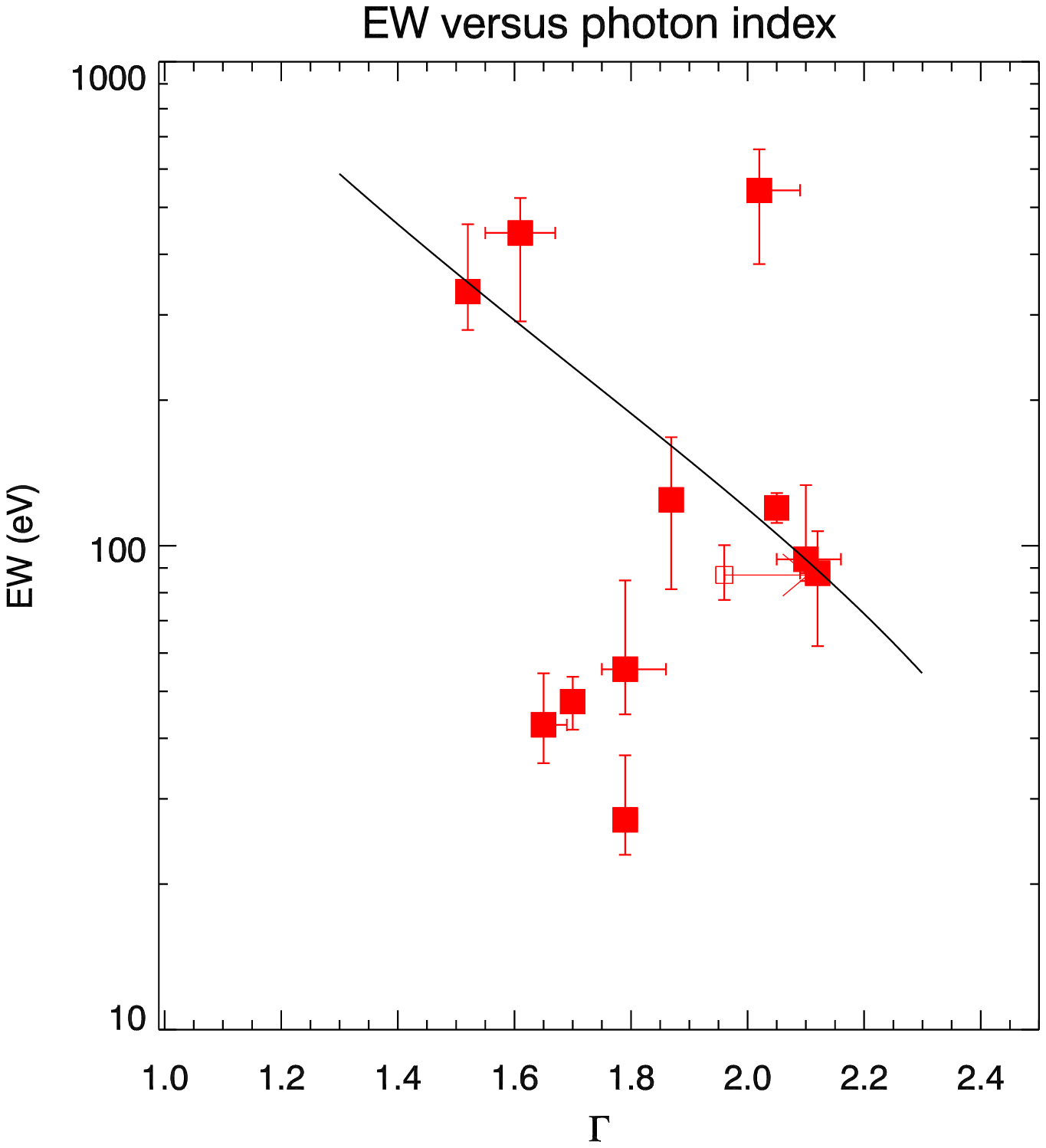}
}
\hbox{
\includegraphics[height=80mm,angle=90]{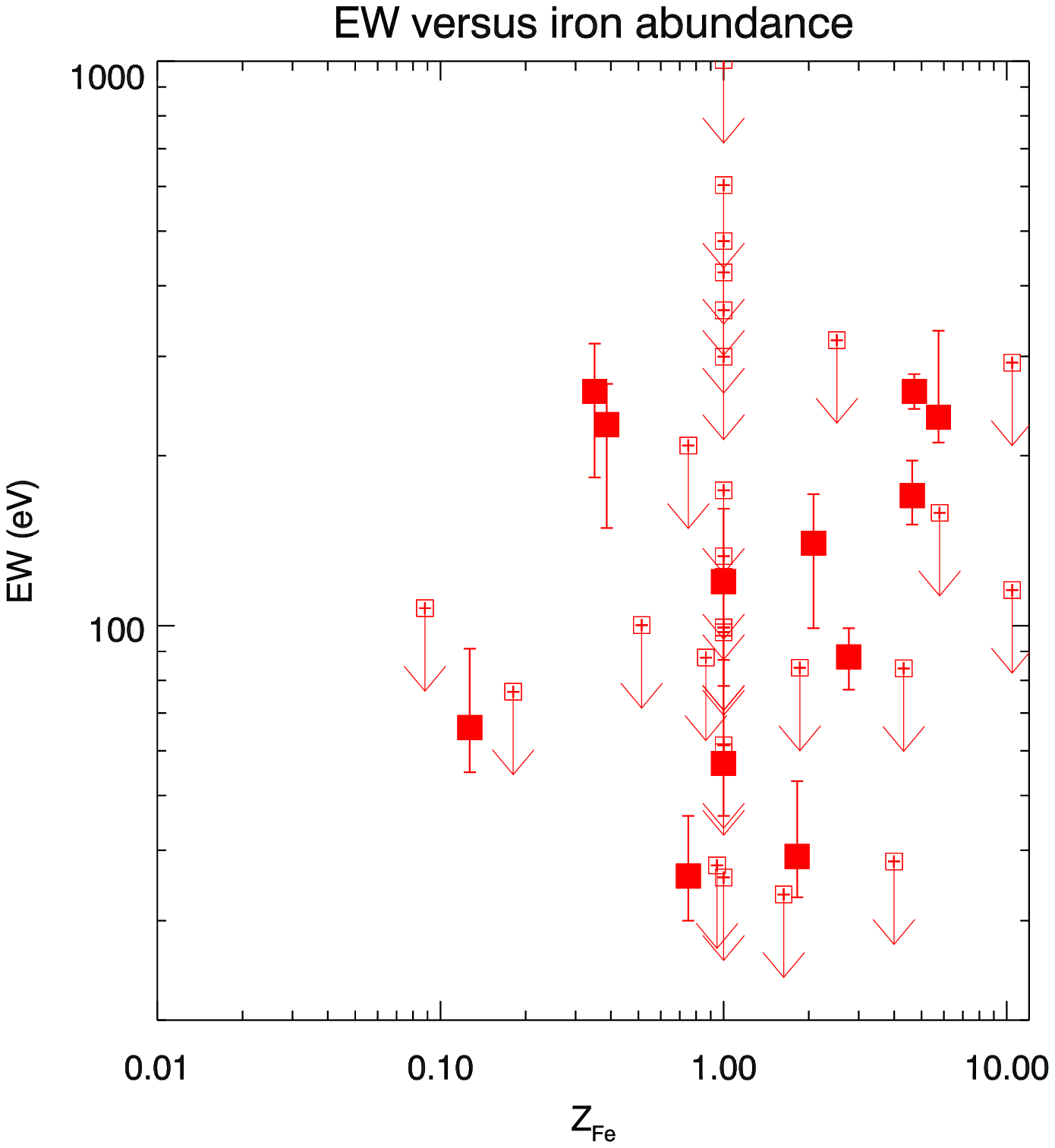}
\hspace{1.0cm}
\includegraphics[height=80mm,angle=90]{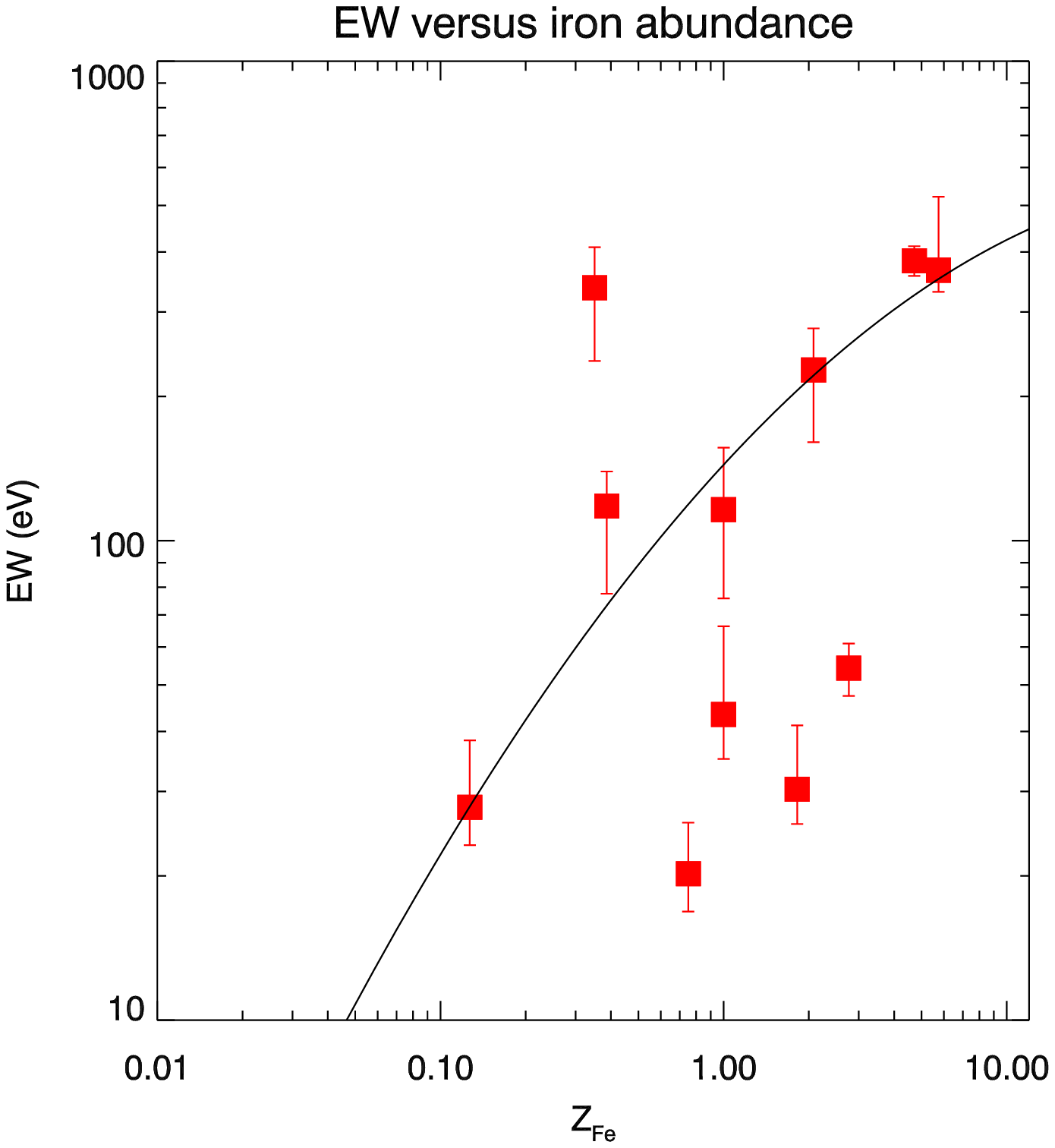}
}
\caption{EW of the relativistically broadened iron line in the FERO+GREDOS sample against the:
disc inclination angle ({\it upper panels}); primary illuminating continuum power-law
spectral index ({\it medium panels}); iron abundance ({\it lower panels}). The
{\it left column} shows the whole sample, the {\it right column} shows only the data points
corresponding to line detections, once the EW values were corrected for the dependency on
the other two parameters according to George \& Fabian (1991) and using the following
common reference values: $\Gamma_0$=1.9;
$\imath_0$=30$^{\circ}$; $Z_{Fe,0}=Z_{\odot}$. The {\it solid line} indicates the
predicted behaviour according to George \& Fabian (1991). {\it Filled symbols}
indicate detections.
}
\label{fig11}
\end{figure*}

In Fig.~\ref{fig11} we show the relation between the EW of the
relativistically broadened iron line and $\Gamma$, $\imath$ and $Z$ for two samples:

\begin{enumerate}

\item the full FERO+GREDOS sample ({\it left panels})

\item the 12 FERO+GREDOS detections only, once the EW values have been corrected for the dependency
of the EW on the other two parameters assuming as common reference the values: $\Gamma_0$=1.9;
$\imath_0$=30$^{\circ}$; $Z_{Fe,0}=Z_{\odot}$.

\end{enumerate}

The EW observed trends against any of the above parameters do not agree with the theoretical
predictions. The correlation against the metallicity
fails to reproduce the whole dynamical range covered by the EW measurements for
$Z_{Fe} \ge 0.5 Z_{\odot}$.

\begin{figure*}
\hbox{
\includegraphics[height=80mm,angle=90]{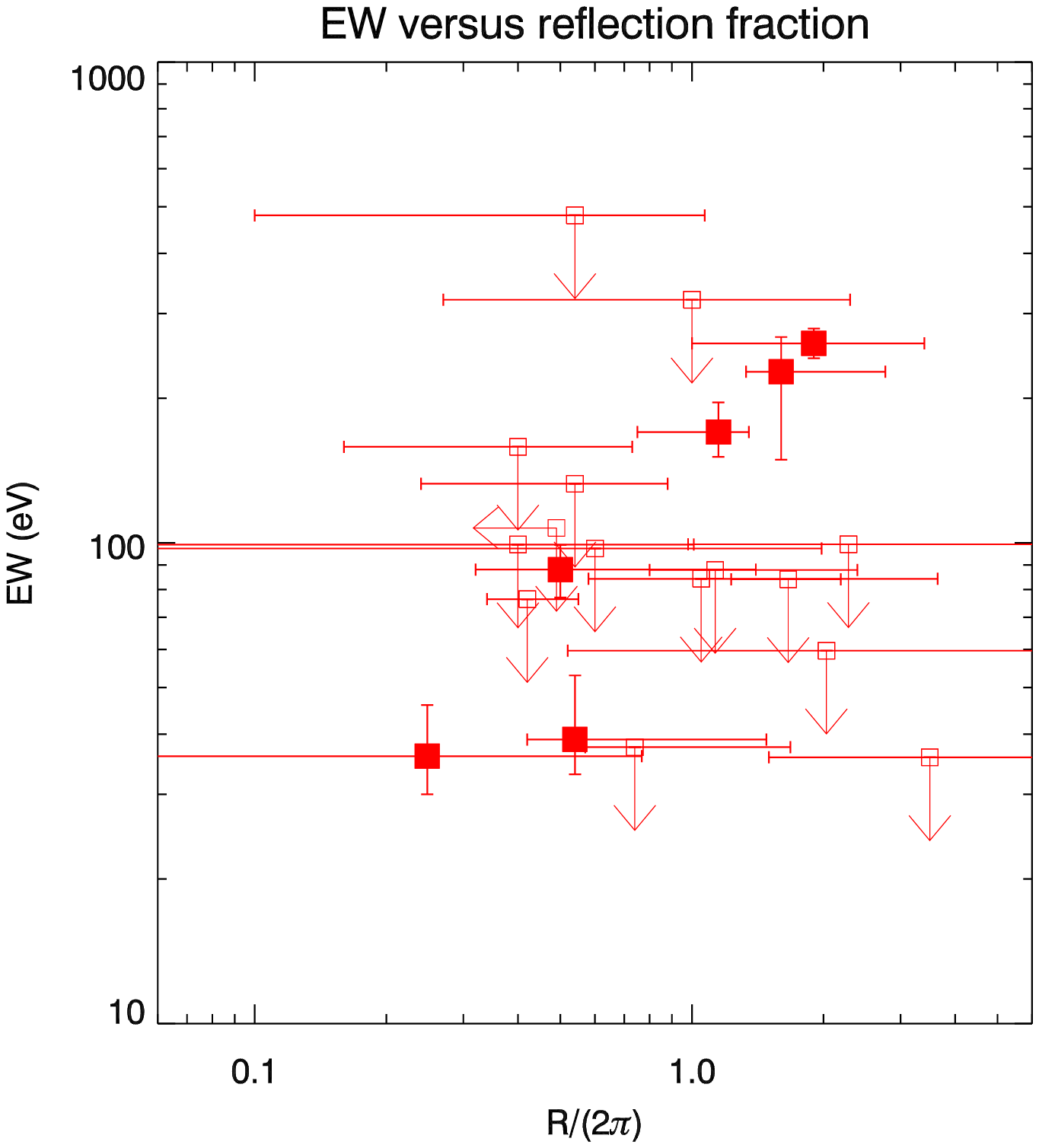}
\hspace{1.0cm}
\includegraphics[height=80mm,angle=90]{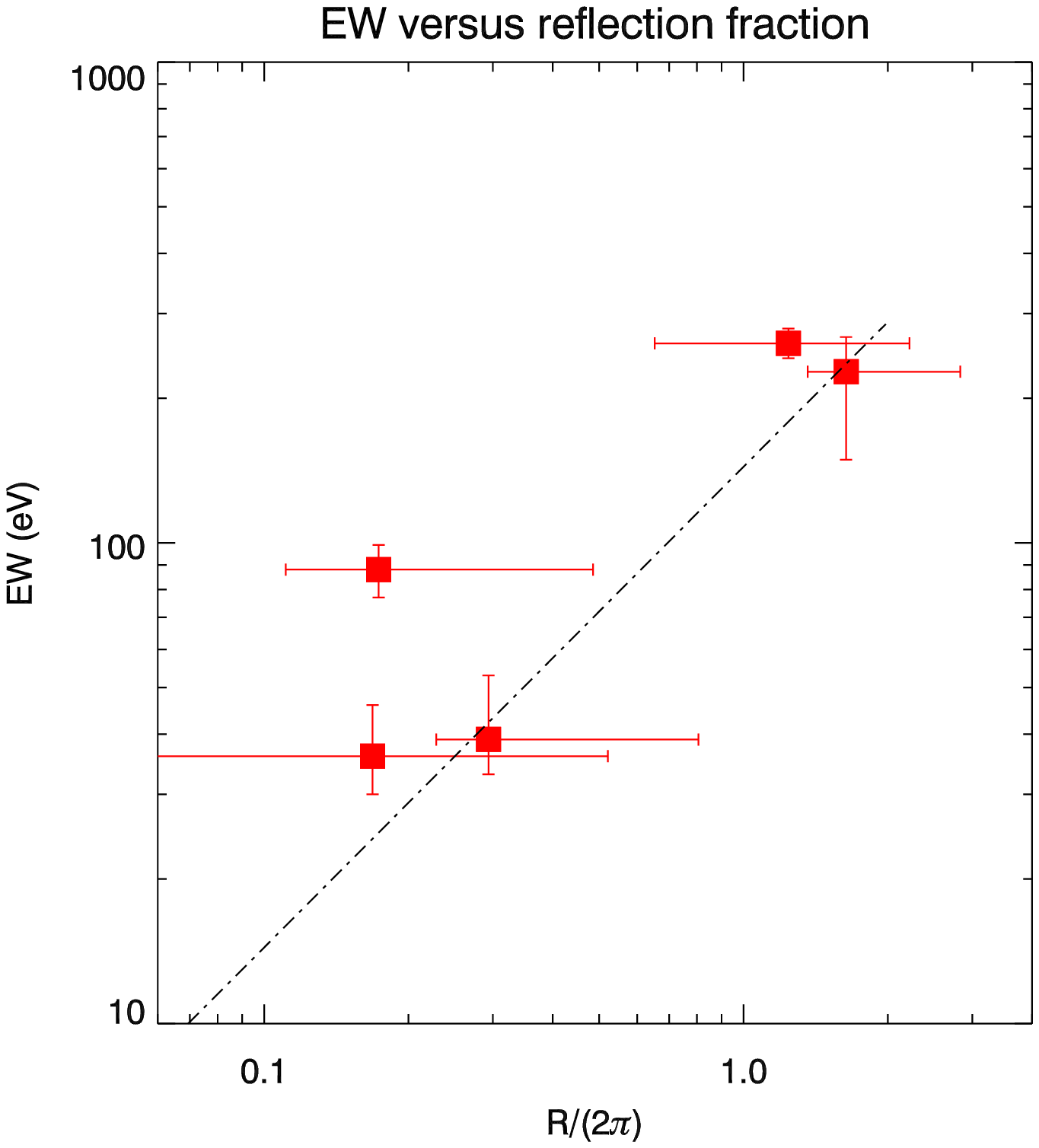}
}
\caption{
EW against the ``reflection fraction'' parameter $R$ in the FERO+GREDOS sample.
{\it Left panel}: all measurements; {\it Right panel}: detections only, corrected
for the EW dependencies on $\Gamma$, $\imath$, and $Z_{Fe}$ to the reference values
of Fig.~\ref{fig11}. The {\it dashed line} indicates a linear relation, with
$EW$=144~eV for $R$=1 (\cite{george91}).
}
\label{fig12}
\end{figure*}

Fig.~\ref{fig11} demonstrates that the falsification tests we aimed at
fails. We are left with the conclusion that the ``reflection fraction'' is the main driver
of the relativistically broadened iron line EW even when X-ray spectra are not disc-reflection
dominated. In Fig.~\ref{fig12} we show the correlation between the EW and the parameter $R$,
which is usually used to characterise quantitatively the reflection fraction. A similar correlation
is found between $\alpha$ and $R$, as expected given the tentative correlation
between $\alpha$ and $EW$ shown in Fig.~\ref{fig8}. $R$
is proportional to the solid angle, $\Omega$, subtended by the disc at the X-ray isotropic source, and
$R=1$ for $\Omega=2\pi$. Unfortunately, the scientific payload on-board XMM-newton does not allow
the simultaneous measurement of $R$. We have therefore derived $R$ from literature 
non-simultaneous measurements. In most cases we have used the results of the BeppoSAX survey
of AGN after \cite{dadina07}, except for those objects for which more recent measurements by
{\it Suzaku} are available: NGC3516 (\cite{markowitz08}), and IC4329A (\cite{markowitz06}).
With these data, the measured values of $R$ is primarily driven by the strength
of the hard ($E \ge 10$~keV) excess above the extrapolation of the primary continuum.
These measurements therefore {\it assume} that the hard X-ray excess is dominated
by reflection. In some cases (\cite{turner09,reeves09}) it appears that absorption may
play a greater role (see also \cite{walton10}).

Care has to be exercised in interpreting this plot, due to the non-simultaneity of the
measurements, as well to the different astrophysical models used
to fit the XMM-Newton/EPIC, the {\it Suzaku} and the BeppoSAX data.
Moreover, a contribution to the continuum Compton-reflection certainly comes from optically
thick matter at pc-scale, responsible for the bulk of the unresolved component of the
narrow K$_{\alpha}$ iron line (\cite{page04,bianchi07,shu10}).
Bearing these caveats in mind, a good
agreement between $R$ and $EW$ is
found, as already reported by previous authors (\cite{perola02,dadina08,walton10}).
Recently, the possibility that a transmitted
component seen through a partial covered absorber could contribute to the emission above 10~keV,
which is crucial for the measurement of the reflection fraction, has been suggested in the
basis of broadband X-ray measurements (\cite{turner09a}). Studies of sizable samples of bright AGN with missions
ensuring a broadband simultaneous coverage such as {\it Suzaku} or the future Astro-H and{\it Astrosat}
would be very valuable in this respect.

\section{Conclusions}

In this paper we have tried to address the question: why does the EW of the relativistically
broadened K$_{\alpha}$ fluorescent iron line in AGN covers a dynamical range of almost two
orders or magnitude if the standard theory (reflection of the X-ray isotropic primary emission
by a plane-parallel infinite accretion disc) predicts variations by at most factor of a few?

In a few cases of very strong ($EW >$300~eV) lines, a ``reflection dominated'' scenario seems
inescapable (\cite{miniutti04}). The advantage of this scenario is that it provides a natural
explanation to a wide range of different phenomenologies commonly observed in X-ray bright
AGN, such as the ``soft excess'' (\cite{crummy06}), and the lack of response of relativistic
reprocessed features to changes of the primary illuminating continuum (\cite{miniutti04,ponti06}).
Does this scenario hold also for ``less extreme'', not reflection dominated AGN? Observationally,
we confirm that the most likely driver for the variation of the EW is the solid angle subtended
by the accretion disc (Fig.~\ref{fig12}).

Resonant trapping of fluorescent photons (\cite{matt93}) can significantly suppress the EW
of line emitted by mildly ionised ($\xi$$\sim$a few hundreds erg~cm~s$^{-1}$. Alternatively,
in many objects of the FERO+GREDOS sample the measured strength of the disk reprocessing
features implies a solid angle subtended by the disk $<$$2 \pi$ (assuming isotropic primary emission).
What is the ultimate physical driver of the lower-than-standard ``reflection fraction'' in weak 
broad-line Seyferts?
One possible way to decrease the ``effective solid angle'' of the reflecting
matter is that the innermost flow is so highly ionised that iron is fully stripped and no
fluorescence or recombination lines cannot occur any longer. The ionisation parameter of the
disc is expected to increase very rapidly close (and beyond) the ISCO (\cite{reynolds08}),
making it quickly ``radiatively truncated''.
The current data are not sensitive to the detailed ionisation state of
the relativistic line profiles. Although this prevents a conclusive answer on this possibility,
there is no compelling evidence so far for a dependence of the line strength on the rest-frame centroid
energy (\cite{delacalle10}).

Already at the dawn of X-ray relativistic spectroscopy, Beloborodov (1999) had invoked bulk motion
of the primary X-ray continuum emitting plasma away from the accretion disc to explain the
weakness of reflection features. Reflection fractions corresponding to $\Omega < \pi$ can be
achieved with moderate ($\beta \equiv v/c \simeq 0.3$) bulk velocities, and
could be further reduced
by scattering of the reflected radiation in an outflowing blob. The X-ray outflowing plasma
could be identified with, {\it e.g.}, the base of a jet, which would at the same time provide
a source of the relativistic electrons required for inverse Compton upscattering of disc photons
into the X-ray regime.

Alternatively, the disc could be physically truncated in sources with a lower-EW line. This mechanism
would also provide a natural explanation to the correlation between
the EW and the power-law index of the disc emissivity as a function of radius (Fig.~\ref{fig8}).
In qualitative terms, a lower EW could be produced by a smaller area of the disc farther away
from the black hole, where the relativistic effects are weaker; this would be reflected in the
way we fit the data by a low ($\alpha \le 2$) values. However, an alternative explanation for the same
correlation invokes a simpler effect related to the quality of the data: when the red wing
of a relativistic line is confused in the underlying continuum, the line profile appears less ``relativistic''
and at the same time weaker. Therefore an observational bias could also produce the correlation in Fig.~\ref{fig8}.
Moreover, claims of truncated discs have been made for sources with very different value of
accretion rate, from low-luminosity AGN to powerful quasars (\cite{matt05,svoboda10,lobban10}).
This is at odds with a similar scenario invoked to explain the hard/low state in Galactic Black
Hole Candidates (GBHCs), typically associated to low accretion states (\cite{fender04}).
A possible way out to reconcile this discrepancy
(apart from invoking a different explanation for the lack of relativistic features in high accretion
rate AGN) is assuming that powerful quasars never go into the equivalent of the GBHC
``low state''. States without relativistic spectral features would correspond to the ``Very High State''
observed, for instance, in the micro-quasar GRS1915+105 (\cite{fender04a}).

The broad iron line could also be produced by localised co-rotating spots on the accretion
disc, which illuminate only a small fraction thereof (\cite{pechacek05}). Indeed, transient
redshifted features have been reported in an handful of objects (\cite{turner02,guainazzi03})
and interpreted in terms of orbiting spots (\cite{dovciak04}). Although doubts have been cast on
the statistical significance of these effects (\cite{vaughan08}), they have been confirmed
by a homogeneous study of a large number of bright sources extracted from the FERO sample
(\cite{demarco09}). The discussion on the reliability of this phenomenology and its 
relevance for the AGN population as a whole is still open, and will be hopefully settled once
more data will be available with future high-throughput missions.

\subsection{On alternatives to the relativistic blurring scenarios}

Finally, if the spectral complexity in the iron band is due to a completely different mechanism,
which does not require general relativistic effects close in the innermost region of the accretion
flow ({\it e.g.}, \cite{turner09}), the observations described in this paper cannot be
explained in terms of relativistic effects. Turner \& Miller (2009) reviewed
recently an alternative scenario where the skewed profile of the K$_{\alpha}$ iron line
is due to partial covering ionised absorption mimicking the broad read wing and the sharp blue
drop at $\simeq$7~keV. This scenario delivers statistically equivalent fits to the
relativistic reflection scenario for a number of time-averaged X-ray spectra of bright
Seyferts (\cite{reeves04,miller09}), as well as explaining their spectral variability
(\cite{miller07b}, 2008, 2010). Readers are referred to
Miller (2007) and Turner \& Miller (2009)
for an exhaustive discussion of strength and weaknesses of both scenarios.
Disk winds are predicted theoretically in several configurations of the accretion flow
(\cite{proga08,sim10}), and their impact on the spectral shape above 2~keV cannot
be neglected (\cite{reeves04}). The authors of the current paper are of the opinion
the not
intuitive geometry that partial covering of the primary X-ray source on scales
of a few gravitation radii (\cite{risaliti07}) requires; and the scale-law
between Galactic Black Hole systems (where partial covering surely does not
occur) and AGN in the timing-domain (\cite{mchardy06}) are serious shortcomings
of the ``partial covering'' scenario. In this context, the possible discovery
of ``clean AGN'' ({\it i.e.}, AGN whose X-ray soft X-ray
spectrum does not show evidence
for a warm absorber; \cite{emmanoulopoulos11}) are a potentially crucial
testbed for the robustness of Fe K$_{\alpha}$ broad line detections.

Turner \& Miller (2009) - and many papers referenced therein - discuss these
shortcomings. They stress that the light bending
scenario - introduced to explain an unexpected lack of correlation between
the short-term variability of the primary continuum and the disk reflection
features - also requires rather ``ad hoc'' assumptions on the X-ray
source geometry and dynamics.
The source has to be compact, and moving in a vertical direction on short
time-scales while maintaining its stability. While the base of a jet
could in principle provide a physical location for this compact, highly
dynamical source of high-energy photons, some of the difficulties of this
picture have been recently discussed by \.Zycki et al. (2010).
In the partial covering scenario the physical scaling of the accreting
system with mass and luminosity implies a scaling of the clumpy covering
material with mass. While short-time variations at high-energy are
still most likely due to intrinsic fluctuations of the source,
the spectral curvature below the Fe K$_{\alpha}$ line and its variation on
long-timescales are successfully modelled with clumpy material in a disk wind.

It remains highly controversial which falsification experiment(s) could
eventually settle the issues. General consensus exists that
studies
of large AGN X-ray high-quality broad-band spectral samples
with operational ({\it Suzaku}; see, {\it e.g.}, the recent work by
Patrick et al., 2010) or
future (Astro-H,{\it Astrosat}, IXO) missions will be crucial.
A statistical comparison of the fit quality produced by different
scenarios on the same sample, as well as of the corresponding observable
distributions, would be a valuable exercise. This paper, the
last of a series of studies inspired by the goal of putting the
discussion on relativistic X-ray spectroscopy
on a sound statistical basis, aims at giving a contribution 
to this ongoing effort.

\section*{Appendix~A. Spectral results on the GREDOS sample}

\begin{figure*}
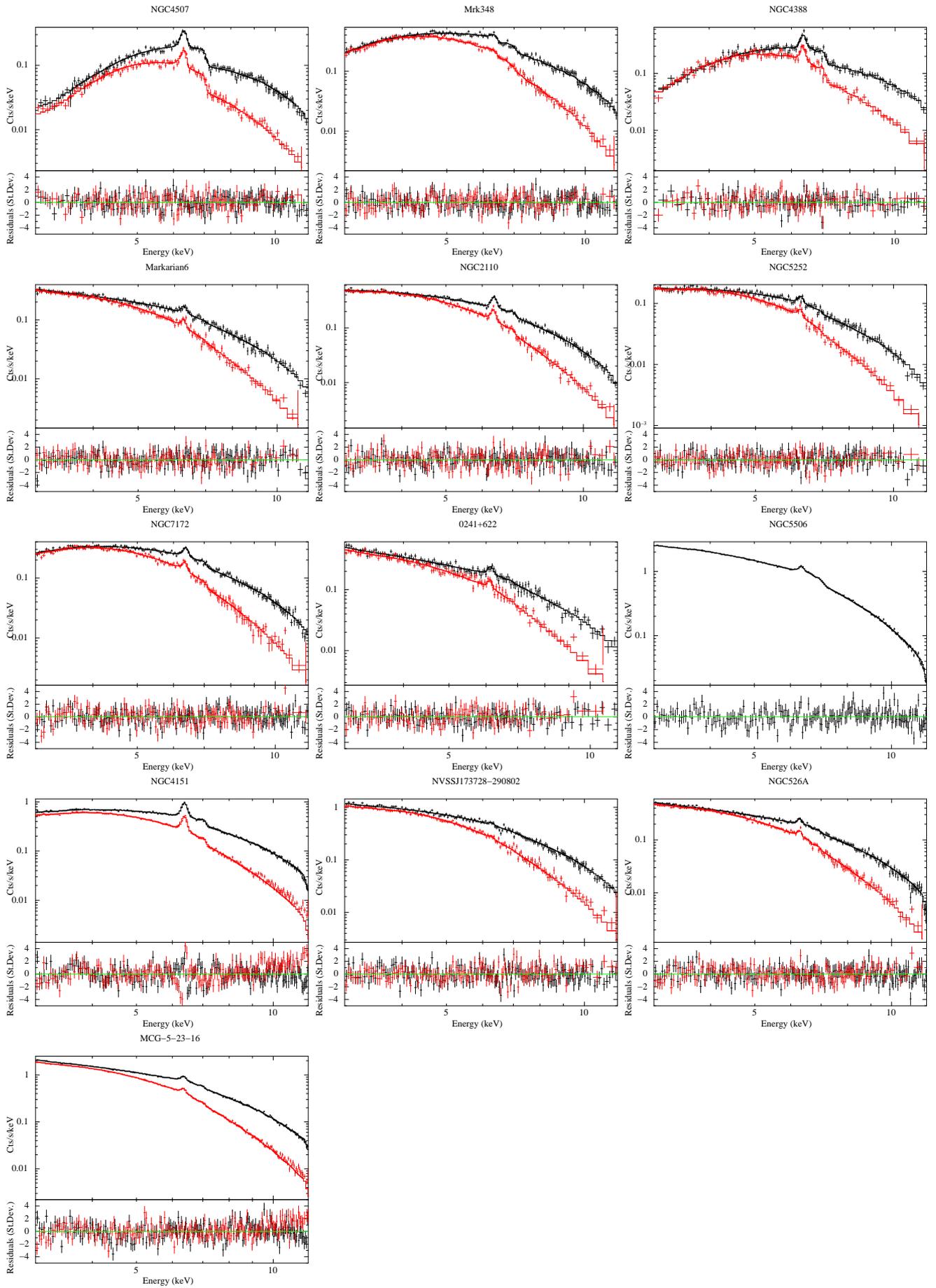

\hbox{
\includegraphics[height=57mm,angle=-90]{fig2a.ps}
\includegraphics[height=57mm,angle=-90]{fig2b.ps}
\includegraphics[height=57mm,angle=-90]{fig2c.ps}
}
\hbox{
\includegraphics[height=57mm,angle=-90]{fig2d.ps}
\includegraphics[height=57mm,angle=-90]{fig2e.ps}
\includegraphics[height=57mm,angle=-90]{fig2f.ps}
}
\hbox{
\includegraphics[height=57mm,angle=-90]{fig2g.ps}
\includegraphics[height=57mm,angle=-90]{fig2h.ps}
\includegraphics[height=57mm,angle=-90]{fig2j.ps}
}
\hbox{
\includegraphics[height=57mm,angle=-90]{fig2k.ps}
\includegraphics[height=57mm,angle=-90]{fig2i.ps}
\includegraphics[height=57mm,angle=-90]{fig2l.ps}
}
\hbox{
\includegraphics[height=57mm,angle=-90]{fig2m.ps}
}
\caption{
Spectra ({\it upper panels}) and residuals in units of standard deviations ({\it lower panels}) when the best-fit
models as in Tab.~\ref{tab3} are applied to the GREDOS sample.
}
\label{fig2}
\end{figure*}

Tab.~\ref{tab3} summarises the results for the continuum components.
\begin{table*}
\caption{Summary of spectral fit results for the GREDOS sample: continuum components}
\label{tab3}
\begin{tabular}{lccccccc} \hline \hline
Source & $\Gamma$ & $N_{H,1}$ & $N_{H,2}$ & $\log(\xi)$/$C_f$$^a$ & $R_{nr}$$^b$ & $R_{r}$$^c$ & Abs.Model$^d$ \\
& & ($10^{22}$~cm$^{-2}$) & ($10^{22}$~cm$^{-2}$) & & & & \\
\hline
NGC4507        & $1.57\pm^{0.07}_{0.14}$ & $58\pm^{41}_5$$d$ & $7 \pm 3$ & $<$2.1 & $0.55 \pm^{0.14}_{0.06}$ & $<0.1$ & CW \\
NGC4388        & $1.86\pm^{0.03}_{0.35}$  & $21.2 \pm^{0.9}_{0.5}$ & $4 \pm 3$ & $2.0 \pm^{0.7}_{0.2}$ & $0.8 \pm 0.3$ & $0.26 \pm^{0.42}_{0.06}$ & CW \\
Mrk6           & $1.43\pm^{0.05}_{0.07}$ & $1.57 \pm^{0.13}_{0.12}$ & ... & $0.85 \pm0.02$ & $<$0.4 & $<$2.9 & PC \\
NGC2110        & $1.69\pm^{0.03}_{0.02}$ & $2.89 \pm^{0.12}_{0.36}$ & $9.3\pm^{1.0}_{2.1}$ & ... & $0.9 \pm 0.5$ & $3.4 \pm 0.7$ & DC \\
NGC5252        & $1.56\pm^{0.32}_{0.19}$ & $3.5\pm^{0.9}_{1.3}$ & $16\pm^{25}_2$ & ... & $<0.1$ & $0.4$ & DC \\
NGC7172        & $1.63\pm^{0.06}_{0.02}$ & $13.3\pm0.2$ & ... & ... & $<0.1$ & $0.68 \pm^{0.10}_{0.03}$ & SC \\
1ES0241+622    & $1.61\pm^{0.09}_{0.04}$ & $0.46\pm^{0.07}_{0.04}$ & ... & ... & $<$0.6 & $<$0.1 & SC \\
GRS1734-292    & $1.41\pm^{0.03}_{0.02}$ & $1.41\pm0.02$ & ... & ... & $<$0.1 & $<$0.4 & SC \\
NGC5506        & $1.689\pm^{0.011}_{0.013}$ & $3.4\pm^{0.7}_{0.5}$ & $1.5\pm^{0.5}_{0.8}$ & $<$-0.27 & $<$0.1 & $<$0.1 & CW \\
NGC4151        & $1.446\pm^{0.010}_{0.009}$ & $13.21\pm^{0.14}_{0.25}$ & ... & $0.853\pm0.004$ & $1.07\pm^{0.09}_{0.56}$ & $<$0.1 & PC \\
NGC526         & $1.457\pm^{0.021}_{0.013}$ & $1.56\pm^{0.04}_{0.03}$ & ... & ... & $<$0.06 & $<$0.12 & SC\\
MCG-05-23-016  & $1.632\pm^{0.010}_{0.016}$ & $1.849\pm^{0.014}_{0.040}$ & ... & ... & $<$0.64 & $<$0.10 & SC \\
\hline \hline
\end{tabular}

\noindent
$^a$ionisation parameter for the warm absorber component for model CW or covering fraction of the patchy absorber for
model PC

\noindent
$^b$ratio between the normalisation of the non-relativistic reflection component and the primary power-law

\noindent
$^c$ratio between the normalisation of the relativistic reflection component and the primary power-law

\noindent
$^d$Absorption model. Legenda: SC~=~single cold absorber; DC~=~double cold absorber; PC~=~partial covering;
CW~=~cold and warm absorber

\end{table*}
In Tab.~\ref{tab4} we report the results on the relativistic line profiles. For non-detected lines we report only the
90\% upper limit on the Equivalent Width for a fixed $E_c \equiv$~6.4~keV.
The parameters refers to a value of the spin corresponding to a maximally
spinning black hole ($a = 0.998$). In all cases, a solution corresponding to a neutral emitting gas
($E_c$=6.4~keV) is preferred.
\begin{table}
\caption{Summary of spectral fit results for the GREDOS sample: relativistic line. All measurements refer to a
rest frame centroid energy $E_c = 6.4$~keV. Upper limits are at the 90\% level.}
\label{tab4}
\begin{tabular}{lcccc} \hline \hline
Source & $\imath$ & $\alpha$ & $EW$ \\
& ($^{\circ}$) & & (eV) \\ \hline
NGC4507 & ... & ... & $<$130 \\                 
Mrk348  & ... & ... & $<$300 \\                  
NGC4388 & ... & ... & ...$^a$ \\                   
Mrk6    & ... & ... & $<$480 \\                        
NGC2110 & ... & ... & $<$100 \\                  
NGC5252 & ... & ... & $<$600 \\                    
NGC7172 & ... & ... & $<$60 \\                   
1ES0241+622  & ... & ... & $<$1000 \\                  
GRS1734-292  & ... & ... & $<$420 \\             
NGC5506  & $40.0\pm^{3.1}_{1.9}$ & $1.88 \pm^{0.16}_{0.23}$ & $120 \pm 40$ \\                 
NGC4151 & ... & ... & $<$80 \\                    
NGC526  & ... & ... & $<$250 \\                    
MCG-05-23-016 & $38\pm^7_5$ &$1.41^{0.04}_{0.30}$ & $55 \pm 26$ \\             
\hline \hline
\end{tabular}

\noindent
$^a$unconstrained

\end{table}
A gallery of spectra and best-fit model is shown in Fig.~\ref{fig2}, whereas the line profiles
(only for detections)
are shown in Fig.~\ref{fig3}. These line profiles are obtained from the data/model ratio against the best-fit model
once the relativistic line profile ({\tt kyrline} component in the best-fit model) is removed.
In a few cases the strength of the continuum
reflection as measured by EPIC is inconsistent with the broad iron line EW, or lack thereof.
We stress that fits of the EPIC-pn spectra are relatively insensitive to the exact value of the reflection
fraction, which is primarily driven by the depth and shape of the iron photo-absorption edge. This feature
is degenerate with the photo-absorption column density in fits of obscured AGN over an energy band limited
to 10~keV. One should therefore refrain from over-interpreting the astrophysical implications of these
results.

\begin{acknowledgements}

Based on observations obtained with XMM-Newton,
an ESA science mission with instruments and contributions directly funded by ESA Member States and NASA.
This research has made use of the NASA/IPAC Extragalactic Database (NED) which is operated by the
Jet Propulsion Laboratory, California Institute of Technology, under contract with the National Aeronautics and
Space Administration. The authors thank G.Cusumano, V.La~Parola and G.Miniutti for comments on an early version of the
manuscript.
Constructive comments by the referee, Lance Miller, substantially improved the quality of the paper, allowing
us to present our results in a more balanced and unbiased way.

\end{acknowledgements}

\end{document}